# Denoising of 3D Magnetic Resonance Images Using a Residual Encoder-Decoder Wasserstein Generative Adversarial Network

Maosong Ran[1], Jinrong Hu[2], Yang Chen[3,4,5], Hu Chen[1], Huaiqiang Sun[6], Jiliu Zhou[1], Yi Zhang[1,7,*]

1. *College of Computer Science, Sichuan University, Chengdu 610065, China*
2. *Department of Computer Science, Chengdu University of Information Technology, Chengdu 610225, China*
3. *Lab of Image Science and Technology, School of Computer Science and Engineering, Southeast University, Nanjing 210096, China*
4. *School of Cyber Science and Engineering, Southeast University, Nanjing 210096, China*
5. *Key Laboratory of Computer Network and Information Integration (Southeast University), Ministry of Education, Nanjing 210096, China*
6. *Department of Radiology, West China Hospital of Sichuan University, Chengdu 610041, China*
7. *Guangdong Provincial Key Laboratory of Medical Image Processing, Southern Medical University, Guangzhou 510515, China*

**Corresponding author: yzhang@scu.edu.cn*

*Abstract*—**Structure-preserved denoising of 3D magnetic resonance imaging (MRI) images is a critical step in medical image analysis. Over the past few years, many algorithms with impressive performances have been proposed. In this paper, inspired by the idea of deep learning, we introduce an MRI denoising method based on the residual encoder-decoder Wasserstein generative adversarial network (RED-WGAN). Specifically, to explore the structure similarity between neighboring slices, a 3D configuration is utilized as the basic processing unit. Residual autoencoders combined with deconvolution operations are introduced into the generator network. Furthermore, to alleviate the oversmoothing shortcoming of the traditional mean squared error (MSE) loss function, the perceptual similarity, which is implemented by calculating the distances in the feature space extracted by a pretrained VGG-19 network, is incorporated with the MSE and adversarial losses to form the new loss function. Extensive experiments are implemented to assess the performance of the proposed method. The experimental results show that the proposed RED-WGAN achieves performance superior to several state-of-the-art methods in both simulated and real clinical data. In particular, our method demonstrates powerful abilities in both noise suppression and structure preservation.**

*Index Items*—**Magnetic Resonance Imaging (MRI); Image Denoising; Deep Learning; Wasserstein GAN; Perceptual Loss.**

## 1 Introduction

Magnetic resonance imaging (MRI) is a noninvasive high-resolution imaging modality that plays a very important role in current clinical diagnostics and scientific research procedures because it can reveal the 3D, internal details and structures of tissues and organs in the human body (Manjon et al.,2008;Mohan et al.,2014;Zhang et al.,2015). However, the quality of MR images can be easily affected by noise during image acquisition, especially when high speed and high resolution are needed. Noise in MR images can not only degrade the imaging quality and the accuracy of clinical diagnoses but also has negative influences on the reliability of subsequent analytic tasks such as registration, segmentation and detection. As a result, efficient algorithms for noise reduction are necessary for further MR analysis.

Over the past few years, a wide variety of postprocessing MRI denoising methods have been developed to improve the imaging quality of MR images (Manjon et al.,2008;Mohan et al.,2014;Zhang et al.,2015). These methods fall into three categories: (a) filtering-based methods (Krissian and Aja-Fernandez,2009;Manjon et al.,2008;Mcveigh et al.,1985;Perona and Malik,1990), (b) domain transform-based methods (Ma and Plonka,2007;Nowak,1999) and (c) statistical methods (Rajan et al.,2012;Sijbers et al.,1998). The filtering-based methods are the most direct and denoise the MR images in the spatial domain. McVeigh et al. (Mcveigh et al.,1985) investigated the results of denoised MR images with both spatial and temporal filters. Most typically, Perona and Malik (Perona and Malik,1990) proposed the classic Perona-Malik (PM) model with a multiscale smoothing and edge detection scheme called an anisotropic diffusion filter. It utilizes gradient information to extract the image structures while reducing the noise. The PM filter and its variants have been successfully extended to 2D and 3D MR images (Gerig et al.,1992;Krissian and Aja-Fernandez,2009;Pal et al.,2017;Samsonov and Johnson,2004). Reducing noise in a transformed domain is different from reducing noise in a spatial domain but is also widely researched, and some typical methods include wavelet and discrete cosine transform (DCT)-based methods (Anand and Sahambi,2009;Hu et al.,2012;Ma and Plonka,2007;Nowak,1999;Pizurica et al.,2003;Yaroslavsky et al.,2001). The statistical approaches first estimate the parameters of Rician noise in noisy MR images. After that, the results are used to yield a statistically optimal denoised image (Awate and Whitaker,2007;Bouhrara et al.,2016;Golshan et al.,2013;He and Greenshields,2009).

Recently, methods based on the self-similarity and sparsity of images have attracted much attention in the field of noise reduction for MRI images. Most algorithms originate from the famous nonlocal means (NLM) filter (Buades et al.,2005), which estimates the current pixel by weighted averaging its similar patches in a search window. One critical drawback of the NLM filter is that it is time consuming. Several variants of NLM filters have been extensively studied to improve this issue for MR Rician denoising (Coupé et al.,2008;Hu et al.,2016;Manjon et al.,2008;Manjón et al.,2012;Wiest-Daesslé et al.,2008). Specifically, in (Coupé et al.,2008), the authors proposed a fast 3D-optimized blockwise version of NLM (ONLM) filter to reduce the computational complexity. To simultaneously make full use of the self-similarity and sparsity of images, the authors in (Manjón et al.,2012) combined the 3D DCT hard-thresholding method and the 3D rotationally invariant version of the nonlocal means filter, which achieved a competitive result. In addition, another state-of-the-art denoising method based on patches is the block-matching and 3D (BM3D) filter (Dabov et al.,2007), which combines the ideas of nonlocality and domain transform (Yaroslavsky et al.,2001). It first groups similar patches into a 3D array, then transforms the array into a frequency domain using DCT or wavelet transform, and finally arrogates multiple estimates at each location. In (Maggioni et al.,2012), the authors developed a filter called BM4D that adapted the BM3D filter to process volumetric data and achieved state-of-the-art performance. Another method similar to BM3D is higher-order singular value decomposition (HOSVD) (Rajwade et al.,2013), which also uses a machine learning technique for noise reduction (Zhang et al.,2015). The difference between BM4D and HOSVD is that the bases of HOSVD are learned from images, which are more adaptive than analytical transforms.

In recent years, the explosive development of deep learning (DL) suggests a kind of new methodology for image processing and computer vision (Girshick,2015;He et al.,2017;He et al.,2016). Except for extensive research on high-level tasks, such as image analysis (Long et al.,2015;Ronneberger et al.,2015), deep learning has been introduced into low-level tasks, including image denoising, deblurring and super-resolution (Dong et al.,2014, 2016;Ledig et al.,2017;Vincent et al.,2010;Zhang et al.,2017). Multilayer perception, autoencoders and convolutional neural networks (CNNs) were used for image restoration and achieved results that were competitive with state-of-the-art methods, such as BM3D, NLM and sparse representation (Zhang et al.,2017). In the field of medical imaging, some pioneering works are given in (Chen et al.,2017a;Chen et al.,2017b;Li and Mueller,2017;Li et al.,2014;Liu and Zhang,2018;Shan et al.,2018;Wang et al.,2016;Xiang et al.,2017;Xu et al.,2017;Yang et al.,2018;Yang et al.,2017a;Yang et al.,2017b;You et al.,2018). However, to the best of our knowledge, the research on MRI denoising is quite limited, and the only work on MRI denoising is represented in (Jiang et al.,2018). The authors proposed a simple, plain CNN for MRI denoising.

Despite extensive research on MRI denoising, current methods suffer from several shortcomings, such as computational burden, nonconvex optimization and/or parameter selection, which seriously impede the practical applications of these methods. In this paper, to conquer these problems and fully explore the potential of the latest techniques in deep learning, we propose an MRI denoising method based on the residual encoder-decoder Wasserstein generative adversarial network (RED-WGAN). The contributions of this paper are fourfold: (a) the proposed model is based on the WGAN framework, which has demonstrated a powerful ability to learn the data distribution in a low dimensional manifold; (b) the ideas of residual networks and autoencoders are utilized to maintain the structural details and edges, which are clinically important; (c) with a proper training procedure, our method yields results that are competitive with several state-of-the-art methods; and (d) our method is highly computationally fast and compatible for parallel implementation on graphic processing units (GPUs).

The rest of this paper is organized as follows. The proposed method is described in Section 2. The experiments and evaluation are given in Section 3. Finally, the results are discussed and conclusions are drawn in Section 4.

## 2 Methods

### 2.1 Noise Reduction Model

One difficulty of MRI denoising is that magnitude images, which are constructed by the real and imaginary parts, are the common form in MRI (Andersen,1996). The noise in magnitude images follows the Rician distribution, which is much more complex than traditional additive noise, such as Gaussian and impulse noise. Many methods were given to statistically model the degradation procedure, and the accuracy of the model heavily affects the final denoising results. DL is an effective way to circumvent this problem, which ignores the physical process and models this procedure corruption by learning from the samples.

The aim of MRI denoising is to recover a high-quality MR image from the corresponding noisy MR image. Let $x \in R^{m\times n}$ denote a noisy MR image and $y \in R^{m\times n}$ denote the corresponding noise-free MR image. The relation between them can be represented as:

$$x = \sigma(y) \tag{1}$$

where $\sigma$ is a mapping function denoting noise contamination. Because the DL-based method is a black box and is independent from the statistical characteristics of the noise, the MR denoising can be simplified to seek the optimal approximation of the function $\sigma^{-1}$, and the denoising procedure can be formulated as:

$$arg\,\min_f \|\hat{y} - y\|_2^2 \tag{2}$$

where $\hat{y} = f(x)$, which is the estimation of $y$, and $f$ denotes the optimal approximation of $\sigma^{-1}$.

### 2.2 Wasserstein GAN

From the viewpoint of statistics, $x$ and $y$ can be regarded as two samples from two different data distributions: noisy image distribution $P_n$ and noise-free image distribution $P_r$, respectively. Then, the denoising operation is a mapping procedure that transforms one distribution to another; that is, the function $f$ maps the samples from $P_n$ to another distribution $P_g$, which is close to $P_r$.

A generative adversarial network (GAN) (Goodfellow et al.,2014) is a kind of generative model that comprises two components: a generative model $G$ and a discriminative model $D$. GANs have been widely applied in many fields, such as image super-resolution (Ledig et al.,2017), image modality transform (Isola et al.,2017) and image generation (Kataoka et al.,2016). The role of the discriminative model is to determine whether a sample is from the generative model distribution $P_g$ or the real data distribution $P_r$, and the generative model generates a new sample from the input sample and tries to make the new sample satisfy the real data distribution $P_r$ as much as possible.

The training process of GANs is a minimax game with the following loss function $L(D,G)$ as

$$\min_G \max_D L(D,G) = E_{y\sim P_r}[\log D(y)] + E_{x\sim P_n}\left[\log\left(1 - D\big(G(x)\big)\right)\right] \tag{3}$$

To solve Eq. (3), $G$ and $D$ are optimized alternatingly.

In (Arjovsky and Bottou,2017), the authors suggested that the training of GAN is difficult because Eq. (3) may lead to a vanishing gradient for the generator $G$ when the discriminator $D$ is fixed. To avoid this problem, an improved variant of the GAN was proposed by Arjovsky and Bottou, called Wasserstein GAN (WGAN) (Arjovsky et al.,2017). Furthermore, Gulrajani et al. presented an improved version of WGAN with a gradient penalty to accelerate the convergence (Gulrajani et al.,2017). The changes in the loss function are as follows:

$$L_{WGAN}(D) = -\mathrm{E}_{y\sim P_r}[D(y)] + \mathrm{E}_{x\sim P_n}\big[D\big(G(x)\big)\big] + \lambda \mathrm{E}_{\hat{x}\sim P_{\hat{x}}}[(\|\nabla_{\hat{x}} D(\hat{x})\|_2 - 1)^2] \tag{4}$$

where the last term is a gradient penalty factor, $\lambda$ is a penalty coefficient, $P_{\hat{x}}$ is a distribution that uniformly samples along straight lines between pairs of points sampled from the real data distribution $P_r$ and the generator distribution $P_g$. The loss function of generator $G$ is formulated as:

$$L_{WGAN}(G) = -\mathrm{E}_{x\sim P_n}\big[D\big(G(x)\big)\big] \tag{5}$$

### 2.3 Combined Loss Function

The MSE loss function is the most common loss function for pixel-level transform tasks, which minimizes the pixelwise differences between the ground truth image and the generated image. It can be calculated as follows:

$$L_{MSE} = \frac{1}{whd}\|G(x) - y\|^2 \tag{6}$$

where $w$, $h$, and $d$ represent the dimensions of the image. Recent studies suggest that although MSE loss function can achieve a high peak signal-to-noise ratio (PSNR), it may suffer from a loss of details, especially high-frequency details, which have a serious impact on clinical diagnostics (Ledig et al.,2017).

To efficiently handle this problem, a perceptual loss is involved in the proposed loss function (Bruna et al.,2015;Gatys et al.,2015;Johnson et al.,2016). A pretrained network can be utilized to extract the features from the ground truth and generated images. The difference between the features from the ground truth image and the generated image is treated as the perceptual similarity. Then, the perceptual loss function is defined as follows:

$$L_{Perceptual} = \frac{1}{whd}\left\|\emptyset\big(G(x)\big) - \emptyset(y)\right\|_F^2 \tag{7}$$

where $\emptyset$ is a feature extractor, and $w$, $h$, and $d$ represent the dimensions of feature maps. In this paper, we apply the pretrained VGG-19 network (Simonyan and Zisserman,2014) to extract the features of the image. The VGG-19 network contains 19 layers: the first 16 layers are convolutional layers, and the subsequent 3 layers are fully connected layers. We only use the first 16 layers as our feature extractor. Then, the specific perceptual loss based on the VGG network is employed as follows:

$$L_{VGG} = \frac{1}{whd}\left\|VGG\big(G(x)\big) - VGG(y)\right\|_F^2 \tag{8}$$

Then, we obtain the weighted joint loss function of generator $G$, which consists of MSE loss, VGG loss and discriminator loss.

$$L_{RED-WGAN} = \lambda_1 L_{MSE} + \lambda_2 L_{VGG} + \lambda_3 L_{WGAN}(G) \tag{9}$$

### 2.4 Network Architectures

The overall architecture of the proposed RED-WGAN network is illustrated in Fig. 1. It consists of a generator network $G$, a discriminator network $D$, and the VGG network is used as the feature extractor. The specific structure of the generator network $G$ is demonstrated in Fig. 2. To accelerate the training procedure and preserve more details, short connections and deconvolution layers are introduced. Furthermore, to explore the ability of the autoencoder to deal with noisy samples, the convolution and deconvolution layers are symmetrically arranged. Specifically, the generator $G$ has an encoder-decoder structure composed of 8 layers: 4 convolutional and 4 deconvolutional layers. Short connections link the corresponding convolution-deconvolutional layer pairs. Except for the last layer, the other layers perform a 3D convolution, a batch-normalization and a LeakyReLU operation in sequence, and the last layer only performs a 3D convolution and a LeakyReLU operation. In this paper, all kernels are set to 3×3×3, and the sequence of the number of filters used is 32, 64, 128, 256, 128, 64, 32, 1.

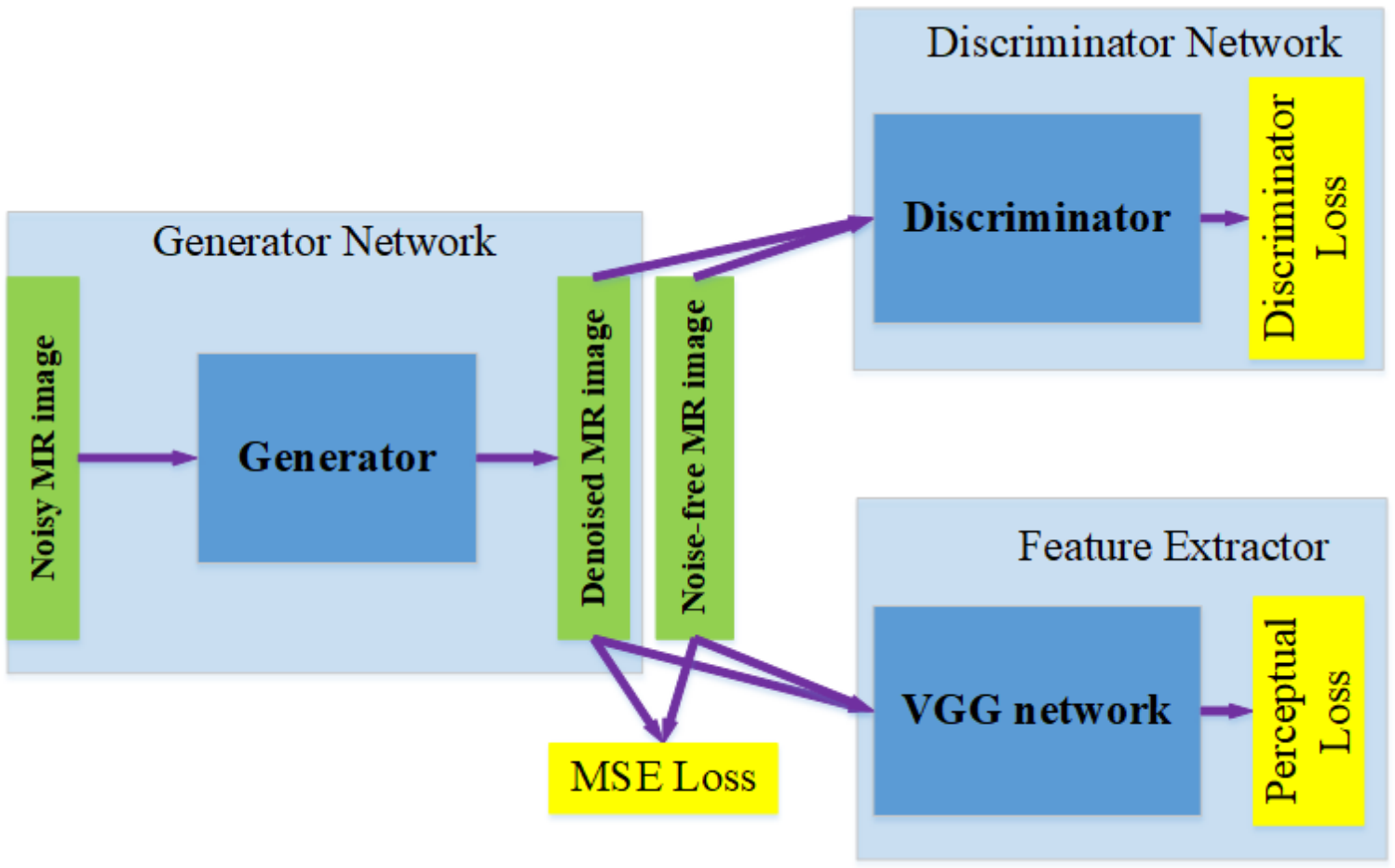

**Fig. 1.** Overall architecture of our proposed RED-WGAN network

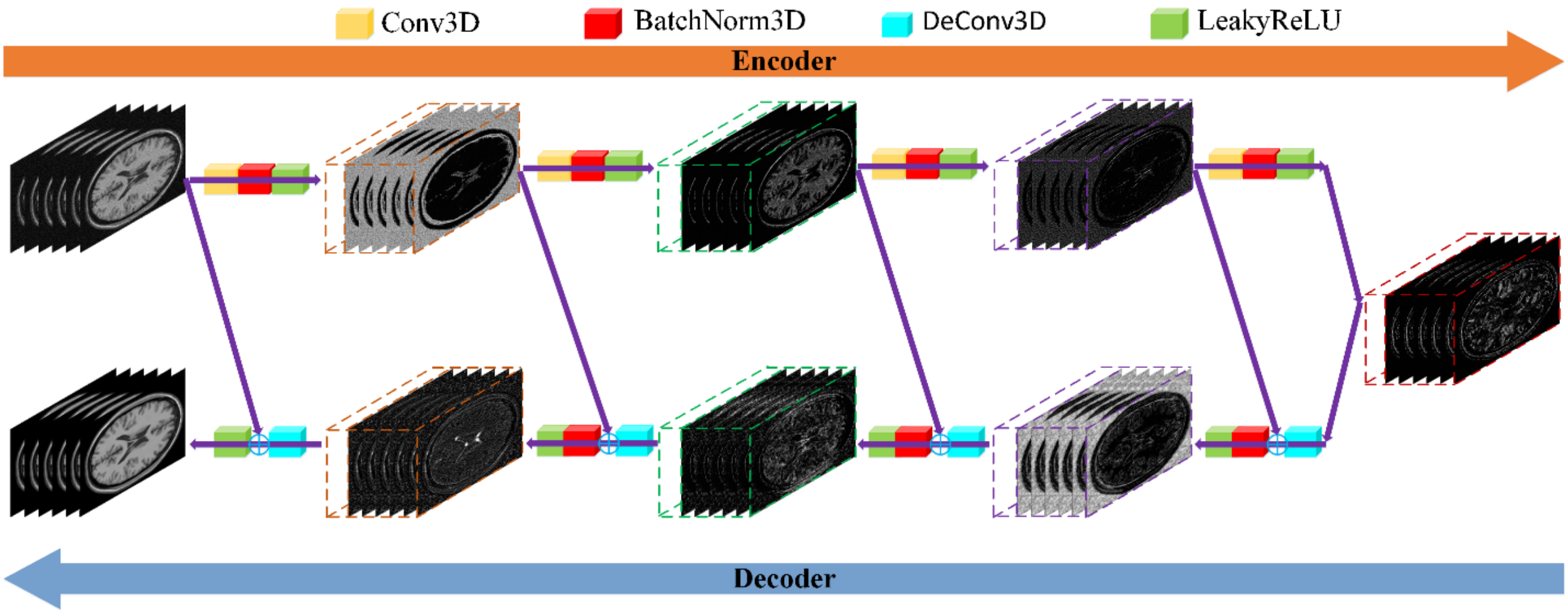

**Fig. 2.** The architecture of the generator network $G$

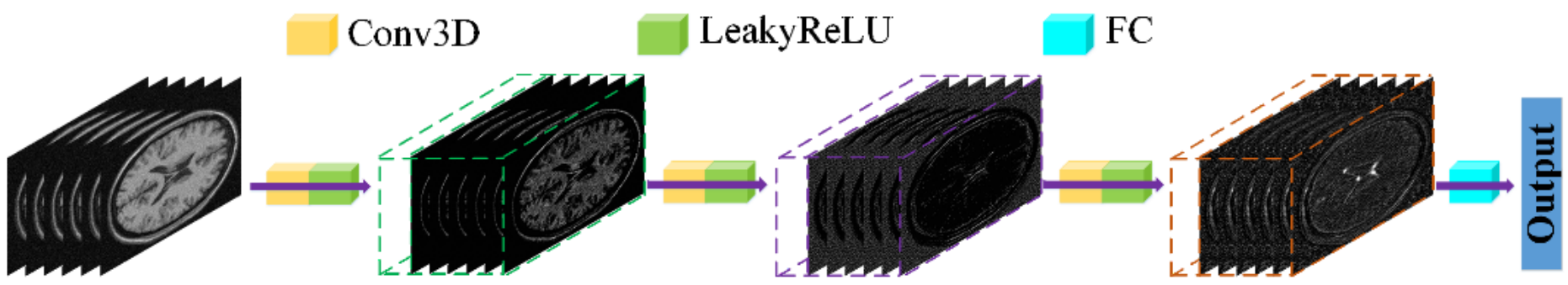

**Fig. 3.** The structure of the discriminator network $D$

The structure of the discriminator network $D$ is illustrated in Fig. 3. It has 3 convolutional layers: one with 32 filters, on with 64 filters and one with 128 filters. The kernel sizes are set to 3×3×3 in all the convolution layers. The last layer is a fully connected layer that has a single output: the discriminant result.

We use a pretrained VGG-19 network to extract the features. For more details, the readers can refer to the original reference (Simonyan and Zisserman,2014). Due to the power of transfer learning (Pan and Yang,2010), there is no need to retrain the network with our target MR images.

## 3 Experiment

### 3.1 Datasets

To validate the performance of the proposed RED-WGAN, extensive experiments on both clinical and simulated datasets were performed.

1) Clinical Data

For the clinical experiments, the well-known IXI dataset (http://brain-development.org/ixi-dataset/), which is collected from 3 different hospitals, was used. The detailed scanning configuration is given in the website mentioned above.

We randomly selected 110 T1-, T2- and PD-weighted brain image volumes from the Hammersmith dataset acquired from a Philips 3T scanner, which is a subset of the IXI dataset. One hundred image volumes were randomly selected as the training set, and the other 10 image volumes from the Hammersmith dataset formed the testing set. To evaluate the robustness of the proposed model for different scanners, 10 image volumes from the Guy's Hospital dataset were also added into the testing set. In the training set, we simulated noisy images by manually adding Rician noise to the images. It is well known that deep learning-based methods require a great deal of training samples, which is very difficult to satisfy, especially in clinics. In this study, to solve this problem, overlapping voxels were extracted from the samples to train the network. This method has been proven efficient in that perceptual differences can be better detected, and the number of samples significantly increases (Dong et al.,2016;ganXie et al.,2012;Jain and Seung,2008). A total of 50000 voxels with of size 32×32×6 are acquired via a fixed sliding step.

2) Simulated Data

For simulated experiments, the BrainWeb database (http://brainweb.bic.mni.mcgill.ca/brainweb/) was used. This dataset contains T1-, T2- and PD-weighted brain images with a size of 181×217×181 with 1×1×1 resolution. Meanwhile, the network trained by the clinical dataset from the Hammersmith Hospital dataset was used to validate the performance and robustness of our model. In the evaluation phase, we chose 6 continuous T1w slices from the middle position of the transverse plane as a test sample to evaluate and compare the performance of the methods.

### 3.2 Training Details

To demonstrate the advantages obtained by our proposed network architecture, two different networks were trained, including RED-WGAN and CNN3D (RED-WGAN with only the generator part and the MSE loss), the latter of which can be seen as an improved version of the method proposed by (Jiang et al.,2018).

Both networks mentioned above were trained on T1-, T2- and PD-weighted brain image volumes with specific noise levels. The parameters $\lambda_1$, $\lambda_2$ and $\lambda_3$ were experimentally set to 1, 0.1 and 1e$^{-3}$, respectively, according to the suggestion in (Ledig et al.,2017;Yang et al.,2017a). Following the suggestions in (Goodfellow et al.,2014), the penalty coefficient $\lambda$ in Eq. (4) was set to 10. The Adam algorithm was used to optimize the loss function (Kinga and Adam,2015), and the parameters for the Adam optimizer were set to $\alpha = 5\mathrm{e}-5, \beta_1 = 0.5, \beta_2 = 0.9$. Our codes for this work are available on https://github.com/Deep-Imaging-Group/RED-WGAN.

### 3.3 Evaluation methods

To validate the performance of the proposed RED-WGAN, three methods (CNN3D, BM4D and PRI-NLM3D (Manjón et al.,2012)) were compared. To evaluate the performance of these methods, three quantitative metrics were employed. The first one is the peak signal-to-noise ratio (PSNR), which considers the root mean square error (RMSE) between the ground truth and denoised images. The second is the structural similarity index measure (SSIM) (Wang et al.,2004), which measures the similarity between ground truth and denoised images. The last one is the information fidelity criterion (IFC) (Hamid Rahim et al.,2005), which quantifies the mutual information between the reference and the testing images to evaluate the perceptual quality.

### 3.4 Results

#### 1) Clinical results

The average quantitative results of BM4D, PRI-NLM3D, CNN3D and RED-WGAN on T1w, T2w and PDw images with different noise levels from 1% to 15% with a step of 2% are illustrated in Tables 1-3. The performances on all metrics of the DL-based methods are significantly superior to traditional denoising algorithms, such as BM4D and PRI-NLM3D. For T1w images, the scores of RED-WGAN are close to CNN3D when the noise level is less than 7%. While the noise level increases, RED-WGAN yields a better performance than the other methods. For T2w images, the results of RED-WGAN are slightly better than all the other methods in most noise levels. In Table 3, the differences are trivial, but the results of CNN3D are slightly better than those of RED-WGAN when the noise level is less than 11%.

**Table 1**

From top to bottom, the PSNR, SSIM and IFC measures of different methods on T1w images with different noise levels

| | 1% | 3% | 5% | 7% | 9% | 11% | 13% | 15% |
|---|---|---|---|---|---|---|---|---|
| | 39.2092 | 29.2209 | 24.6349 | 21.6248 | 19.3978 | 17.6280 | 16.1530 | 14.8845 |
| Noise | 0.8325 | 0.6007 | 0.4964 | 0.4242 | 0.3667 | 0.3186 | 0.2771 | 0.2417 |
| | 6.9618 | 3.8510 | 2.7399 | 2.1268 | 1.7276 | 1.4405 | 1.2262 | 1.0577 |
| | 43.7217 | 37.3037 | 34.5095 | 32.6762 | 31.3338 | 29.7973 | 28.1597 | 25.9018 |
| BM4D | 0.9832 | 0.9393 | 0.9034 | 0.8926 | 0.8798 | 0.8622 | 0.8417 | 0.8126 |
| | 7.3469 | 4.6026 | 3.6226 | 3.0635 | 2.6889 | 2.3640 | 2.0850 | 1.8054 |
| | 42.5101 | 36.7709 | 33.8254 | 31.3052 | 29.4420 | 27.9812 | 26.8905 | 26.3974 |
| PRI-NLM3D | 0.9601 | 0.9357 | 0.8854 | 0.7830 | 0.7432 | 0.6767 | 0.6816 | 0.6664 |
| | 6.9492 | 4.4683 | 3.4591 | 2.8532 | 2.4492 | 2.1196 | 1.8879 | 1.6662 |
| | **44.7101** | **38.4564** | **35.8638** | **33.6071** | 32.7940 | 31.4896 | 29.9069 | 28.6901 |
| CNN3D | **0.9867** | **0.9542** | **0.9293** | **0.9091** | 0.9005 | **0.8927** | 0.8659 | 0.8525 |
| | **7.6368** | **4.9561** | **4.0391** | 3.2551 | 3.0046 | 2.6684 | 2.1961 | 2.0109 |
| | 44.4336 | 36.5281 | 34.4664 | 33.0387 | **33.0367** | **32.1459** | **30.5995** | **29.5566** |
| RED-WGAN | 0.9806 | 0.9205 | 0.8957 | 0.8957 | **0.9021** | **0.8927** | **0.8779** | **0.8679** |
| | 7.5411 | 4.5158 | 3.7543 | **3.2673** | **3.0780** | **2.8640** | **2.4169** | **2.1702** |

**Table 2**

From top to bottom, the PSNR, SSIM and IFC measures of different methods on T2w images with different noise levels.

| | 1% | 3% | 5% | 7% | 9% | 11% | 13% | 15% |
|---|---|---|---|---|---|---|---|---|
| | 39.4202 | 29.3326 | 24.7211 | 21.7165 | 19.4976 | 17.7243 | 16.2557 | 14.9919 |
| Noise | 0.8608 | 0.6026 | 0.4751 | 0.3915 | 0.3304 | 0.2816 | 0.2417 | 0.2093 |
| | 7.3619 | 3.9930 | 2.8271 | 2.1905 | 1.7797 | 1.4861 | 1.2607 | 1.0919 |
| | 44.0309 | 37.7481 | 34.8165 | 32.5272 | 30.1290 | 28.0781 | 24.6566 | 20.7909 |
| BM4D | 0.9758 | 0.9238 | 0.8856 | 0.8582 | 0.8222 | 0.7935 | 0.7315 | 0.6267 |
| | 7.7125 | 4.7985 | 3.7447 | 3.0918 | 2.6203 | 2.2847 | 1.9444 | 1.6280 |
| | 43.1845 | 37.2417 | 33.9030 | 31.3343 | 30.0204 | 28.1753 | 26.4021 | 26.0070 |
| PRI-NLM3D | 0.9615 | **0.9320** | 0.8494 | 0.7550 | 0.7380 | 0.6749 | 0.6005 | 0.6042 |
| | 7.4173 | **4.6000** | 3.5335 | 2.9057 | 2.4820 | 2.1621 | 1.9017 | 1.6926 |
| | 44.9166 | 38.5515 | 36.2237 | 34.4047 | 33.0077 | 31.9171 | 30.7051 | **29.9941** |
| CNN3D | 0.9757 | 0.9210 | 0.8972 | 0.8833 | 0.8657 | 0.8511 | 0.8274 | **0.8197** |
| | 7.8464 | 5.0305 | 4.1207 | 3.5154 | 3.0997 | 2.8026 | 2.4815 | **2.2502** |
| | **44.9592** | **38.5799** | **36.2267** | **34.5710** | **33.0959** | **31.9171** | **30.7556** | 29.7998 |
| RED-WGAN | **0.9769** | 0.9223 | **0.8972** | **0.8838** | **0.8672** | **0.8511** | **0.8294** | 0.8181 |
| | **7.8499** | 5.0346 | **4.1170** | **3.5556** | **3.1483** | **2.8026** | **2.4974** | 2.2272 |

**Table 3**

From top to bottom, the PSNR, SSIM and IFC measures of different methods on PDw images with different noise levels

| | 1% | 3% | 5% | 7% | 9% | 11% | 13% | 15% |
|---|---|---|---|---|---|---|---|---|
| | 39.3873 | 29.3073 | 24.6963 | 21.6903 | 19.6796 | 17.6862 | 16.2160 | 14.9550 |
| Noise | 0.8536 | 0.5840 | 0.4510 | 0.3668 | 0.3128 | 0.2620 | 0.2262 | 0.1970 |
| | 6.7055 | 3.6066 | 2.5488 | 1.9745 | 1.6429 | 1.3399 | 1.1451 | 0.9948 |
| | 44.7089 | 38.6318 | 36.0036 | 34.2296 | 32.6456 | 30.9293 | 29.0103 | 26.3797 |
| BM4D | 0.9787 | 0.9266 | 0.8898 | 0.8703 | 0.8547 | 0.8342 | 0.8124 | 0.7857 |
| | 7.1283 | 4.4597 | 3.5642 | 3.0383 | 2.6771 | 2.3371 | 2.0549 | 1.8061 |
| | 43.8139 | 37.6367 | 34.5373 | 32.4614 | 31.2401 | 29.3602 | 28.3420 | 27.333 |
| PRI-NLM3D | 0.9721 | 0.9058 | 0.8520 | 0.8131 | 0.7900 | 0.7250 | 0.6946 | 0.6606 |
| | 6.8131 | 4.2496 | 3.2572 | 2.6843 | 2.3273 | 1.9870 | 1.7431 | 1.6410 |
| | 45.7124 | 39.8149 | **37.1052** | **34.9915** | **33.6963** | **33.0781** | **31.5944** | 30.5857 |
| CNN3D | **0.9839** | **0.9447** | **0.8966** | **0.8739** | **0.8559** | **0.8549** | **0.8319** | 0.8037 |
| | **7.2873** | **4.7514** | **3.8276** | **3.2373** | **2.8932** | **2.6220** | 2.2493 | 2.1192 |
| | **45.7125** | **39.8201** | 37.1031 | 34.9867 | 33.6870 | 33.0561 | 31.3285 | **31.5303** |
| RED-WGAN | 0.9836 | 0.9452 | 0.8965 | 0.8738 | 0.8556 | 0.8546 | 0.8154 | **0.8197** |
| | 7.2848 | 4.7519 | 3.8251 | 3.2014 | 2.8747 | 2.6078 | **2.2788** | **2.3422** |

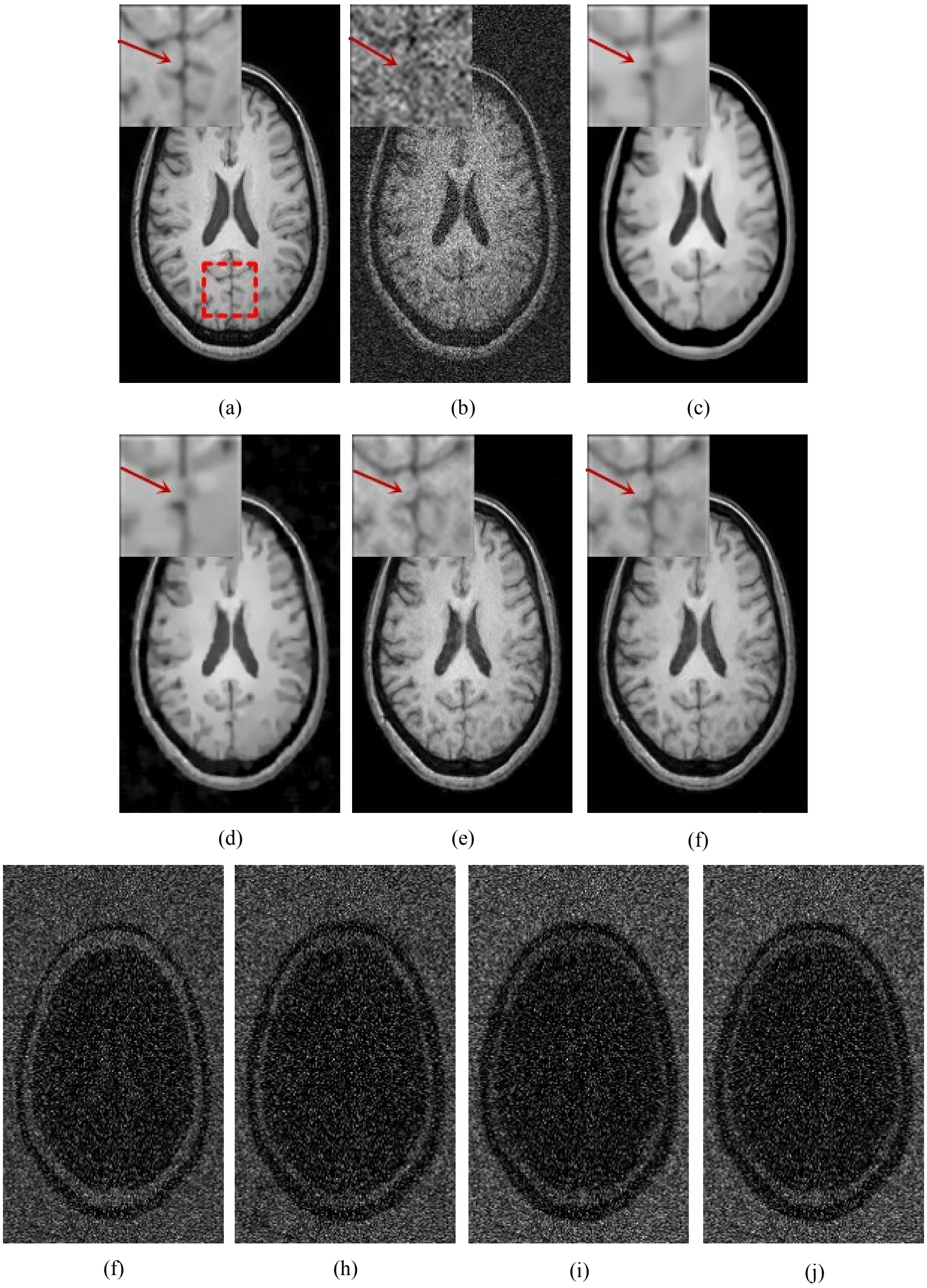

**Fig. 4.** One denoised T1w example from the testing set with 15% Rician noise. (a) Noise-free image, (b) Noisy image, (c) BM4D, (d) PRI-NLM3D, (e) CNN3D, (f) RED-WGAN, (g) Residual of BM4D, (h) Residual of PRI-NLM3D, (i) Residual of CNN3D, (j) Residual of RED-WGAN.

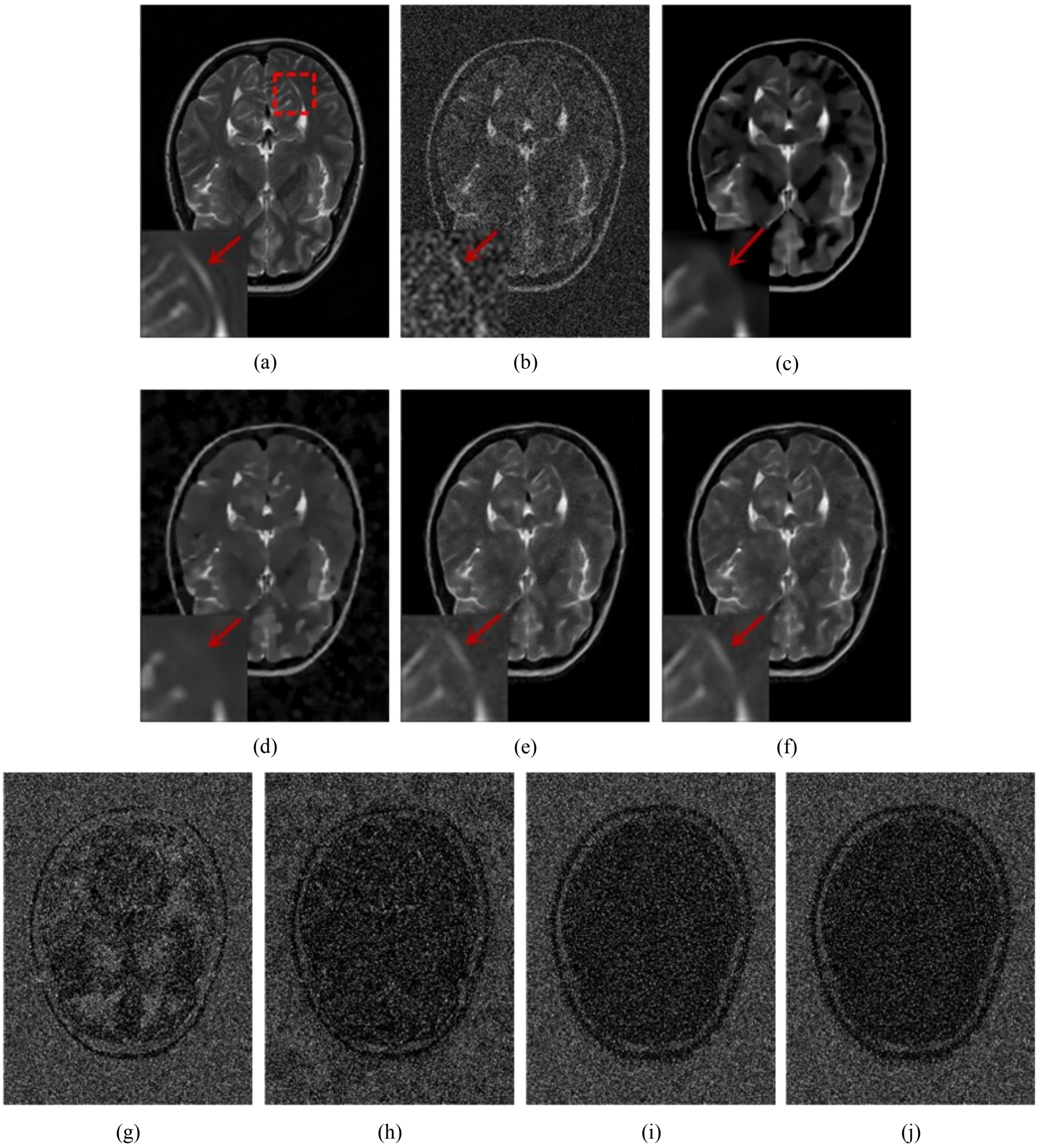

**Fig. 5.** One denoised T2w example from the testing set with 15% Rician noise. (a) Noise-free image, (b) Noisy image, (c) BM4D, (d) PRI-NLM3D, (e) CNN3D, (f) RED-WGAN, (g) Residual of BM4D, (h) Residual of PRI-NLM3D, (i) Residual of CNN3D, (j) Residual of RED-WGAN.

**Table 4**

Quantitative results associated with different methods for Figs. 4(T1w), 5(T2w) and 6(PDw).

| Method | T1w | | | T2w | | | PDw | | | Average Execution Time |
|---|---|---|---|---|---|---|---|---|---|---|
| | PSNR | SSIM | IFC | PSNR | SSIM | IFC | PSNR | SSIM | IFC | |
| Noise | 14.7437 | 0.2603 | 1.0887 | 14.6613 | 0.1797 | 0.9352 | 14.2264 | 0.2164 | 1.0564 | |
| BM4D | 27.0763 | 0.8354 | 2.1528 | 22.9486 | 0.7173 | 1.5034 | 25.7049 | 0.7776 | 1.7739 | 5.73 |
| PRI-NLM3D | 28.2652 | 0.7901 | 2.0454 | 26.0788 | 0.5612 | 1.2938 | 26.3631 | 0.6897 | 1.4646 | 4.16 |
| CNN3D | 29.1561 | 0.8742 | 2.2831 | 29.8515 | 0.8254 | 1.9280 | 28.3651 | 0.7893 | 2.1017 | 0.17 |
| RED-WGAN | **30.0584** | **0.8892** | **2.4639** | **29.9575** | **0.8351** | **1.9467** | **31.1590** | **0.8624** | **2.6937** | **0.16** |

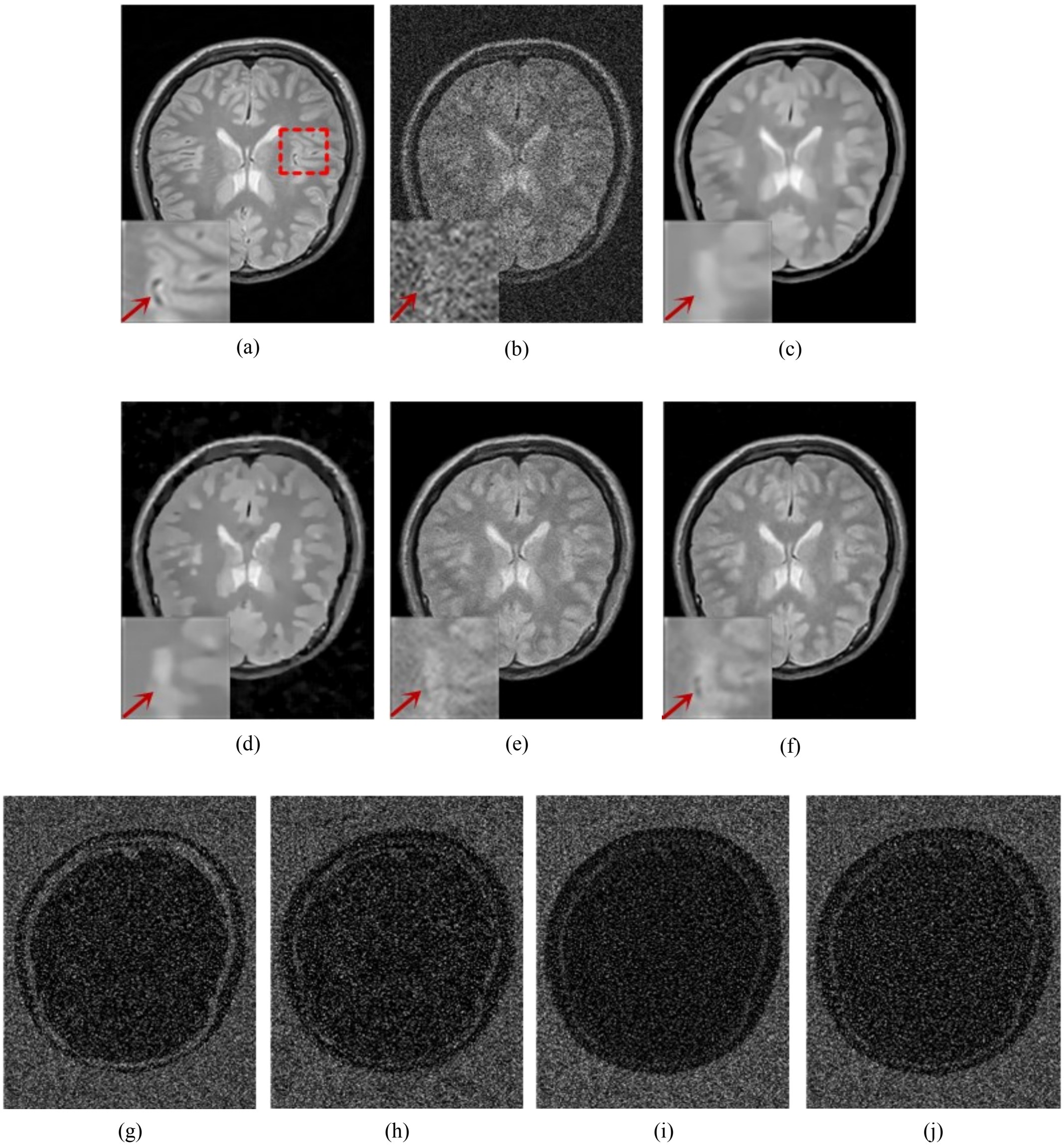

**Fig. 6.** One denoised PDw example from the testing set with 15% Rician noise. (a) Noise-free image, (b) Noisy image, (c) BM4D, (d) PRI-NLM3D, (e) CNN3D, (f) RED-WGAN, (g) Residual of BM4D, (h) Residual of PRI-NLM3D, (i) Residual of CNN3D, (j) Residual of RED-WGAN.

Figs. 4-6 provide a visual evaluation of the different results for T1w, T2w and PDw brain images selected from the Hammersmith Hospital dataset in the testing set and corrupted with 15% Rician noise. All of the methods can suppress noise to varying degrees. However, BM4D and PRI-NLM3D suffer from obvious oversmoothing effects and distort some important details, which can be better sensed in Figs. 5 and 6. Both RED-WGAN and CNN3D efficiently avoid oversmoothing and preserve more structural details than BM4D and PRI-NLM3D. RED-WGAN outperforms CNN3D in noise suppression and obtains the most consistent results with respect to the reference images. The quantitative

results from different methods for Figs. 4-6 are summarized in Table 4. The results suggest that RED-WGAN achieved the best performance in terms of PSNR, SSIM and IFC on all the modalities, which is consistent with the visual inspection. Although the quantitative results of RED-WGAN and CNN3D are similar, the visual effect of RED-WGAN is significantly better than CNN3D in the enlarged regions. This benefit is due to the introduction of WGAN and the combined loss function, which can efficiently generate results that are closer to the original data distribution. In Table 4, we also demonstrate the running times for different methods. It is clear that CNN3D and our proposed RED-WGAN are much faster than other traditional methods. Once the deep learning-based method finishes training, forward propagation is very fast.

To further demonstrate the robustness of the proposed RED-WGAN approach, the Guy's Hospital dataset, which is another subset of the IXI dataset, was included as the testing set. Figs. 7-9 show a denoised example in different modalities with different methods on the Guy's Hospital dataset with 15% Rician noise. Table 5 summarizes the corresponding quantitative results. Although CNN3D and RED-WGAN were trained with a different training set, which has different scanning parameters to the testing set, we still obtained satisfactory results in Figs. 7-9. Most of the noise is efficiently removed, and the structural details are better preserved in the results from RED-WGAN. Some ROIs, indicated by red dotted boxes, were magnified to further demonstrate the differences produced by different methods. It can be noted that RED-WGAN maintains the details better than other methods, which lose the details to varying degrees. Furthermore, DL-based methods demonstrate a robust performance for the datasets obtained with different scanners and scanning parameters.

**2）Simulated results**

One representative T1w result with 9% Rician noise from the BrainWeb dataset is shown in Fig. 10. In Fig. 10, all the methods can eliminate most of the noise, but in the residual images, BM4D lost part of the structural details. From the enlarged regions, it is clear that PRI-NLM3D and DL-based methods perform better than BM4D in terms of structure preservation. Meanwhile, RED-WGAN obtains the best scores in all the metrics, even though the model was trained with clinical data.

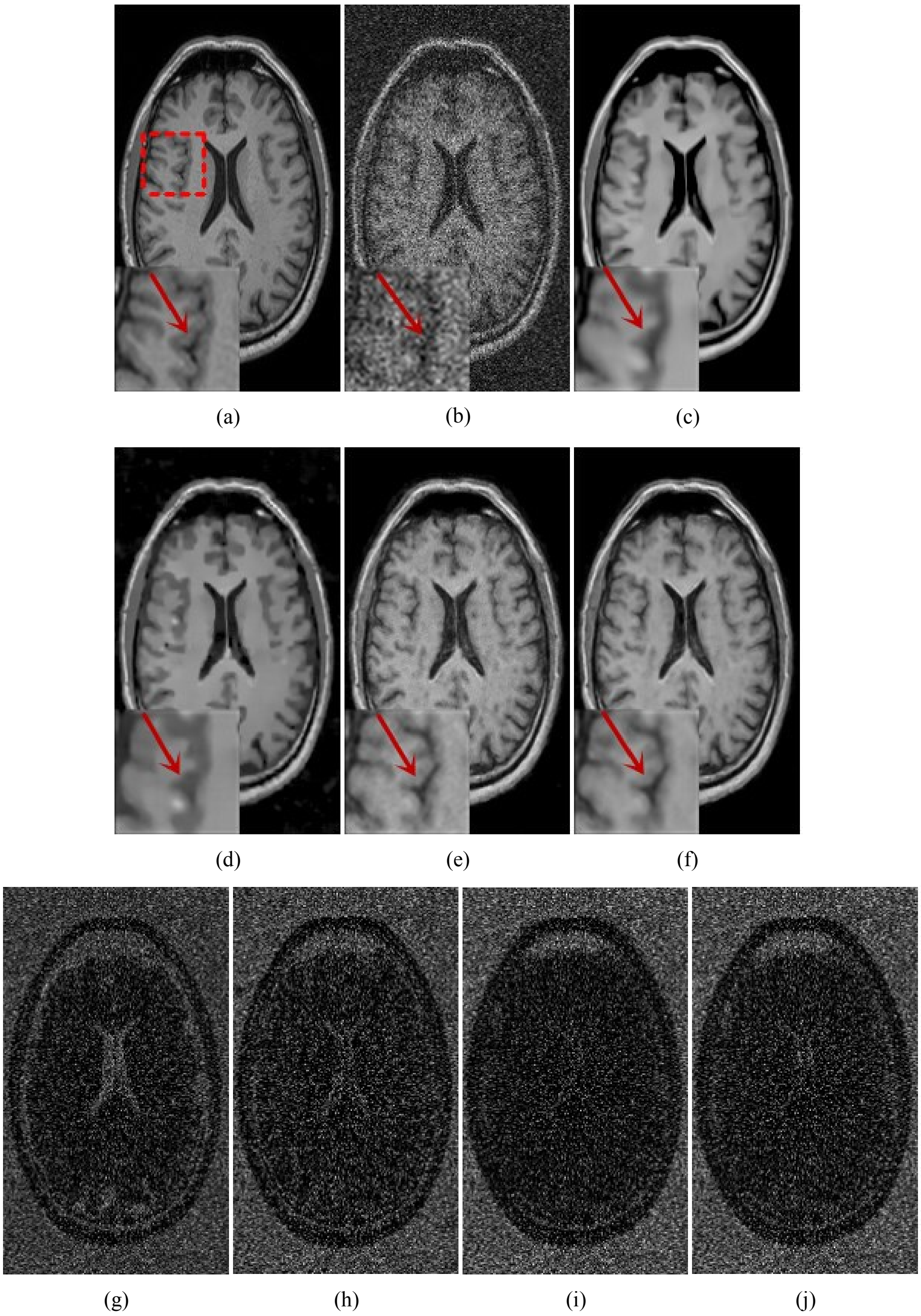

Fig. 7. One denoised T1w example from the Guy's Hospital dataset with 15% Rician noise. (a) Noise-free image, (b) Noisy image, (c) BM4D, (d) PRI-NLM3D, (e) CNN3D, (f) RED-WGAN, (g) Residual of BM4D, (h) Residual of PRI-NLM3D, (i) Residual of CNN3D, (j) Residual of RED-WGAN

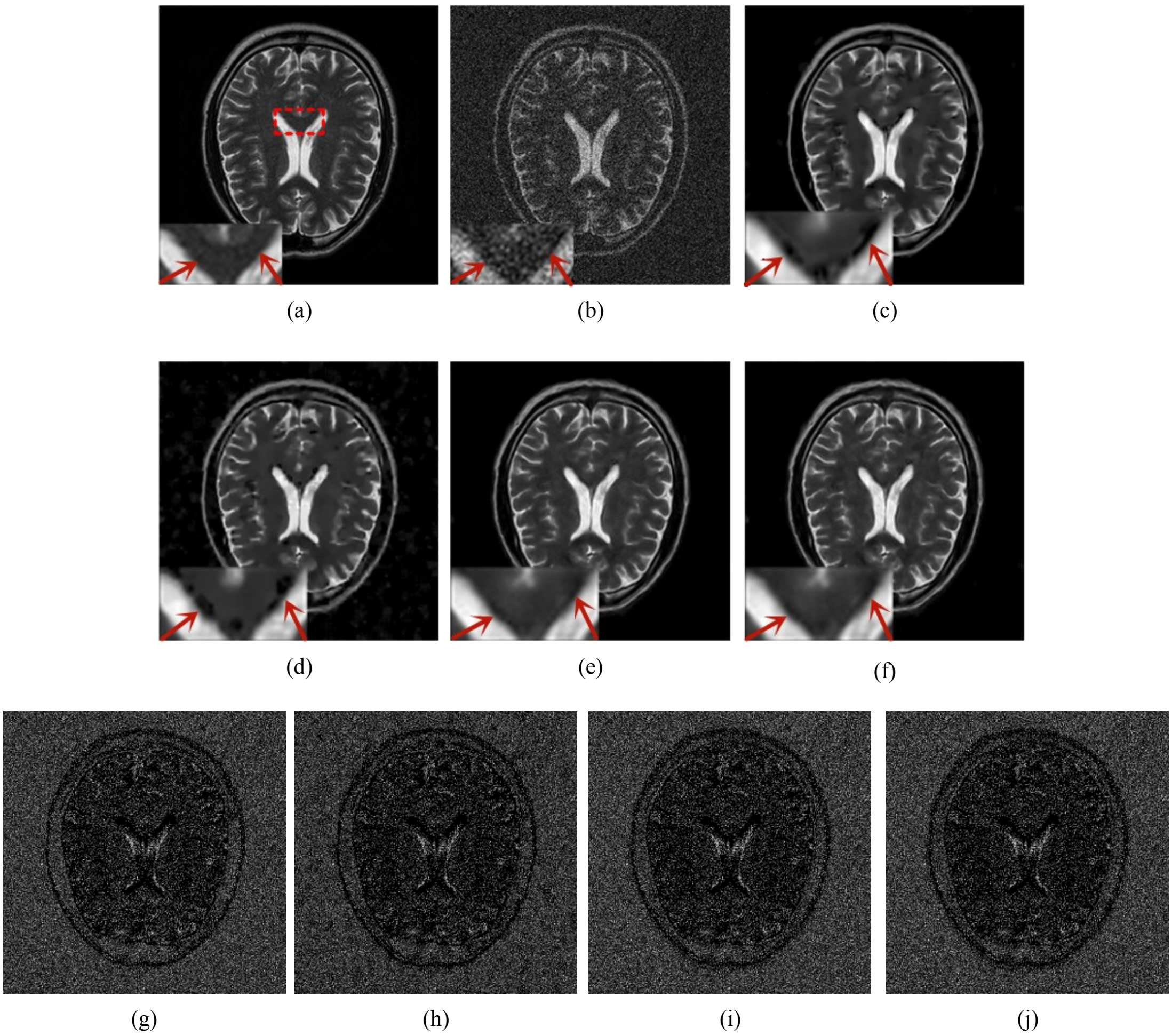

**Fig. 8.** One denoised T2w example from the Guy's Hospital dataset with 15% Rician noise. (a) Noise-free image, (b) Noisy image, (c) BM4D, (d) PRI-NLM3D, (e) CNN3D, (f) RED-WGAN, (g) Residual of BM4D, (h) Residual of PRI-NLM3D, (i) Residual of CNN3D, (j) Residual of RED-WGAN.

**Table 5**

Quantitative results associated with different method outputs for Figs. 7 (T1w), 8 (T2w) and 9 (PDw)

| Method | T1w | | | T2w | | | PDw | | |
|---|---|---|---|---|---|---|---|---|---|
| | PSNR | SSIM | IFC | PSNR | SSIM | IFC | PSNR | SSIM | IFC |
| Noise | 14.5527 | 0.3020 | 1.2473 | 14.2563 | 0.2183 | 1.1340 | 14.1508 | 0.1874 | 1.0456 |
| BM4D | 24.3476 | 0.8279 | 2.1817 | 24.4251 | 0.7949 | 1.9152 | 27.5299 | 0.8525 | 1.5781 |
| PRI-NLM3D | 25.1755 | 0.7701 | 1.7740 | 23.5693 | 0.6480 | 1.4618 | 28.5738 | 0.6984 | 1.3268 |
| CNN3D | 26.2868 | 0.8372 | 2.1055 | 25.1037 | **0.8211** | **2.1484** | **30.4959** | **0.8611** | 1.9326 |
| RED-WGAN | **27.1845** | **0.8645** | **2.2909** | **25.0004** | 0.8174 | 2.1295 | 30.2283 | 0.8323 | **1.9430** |

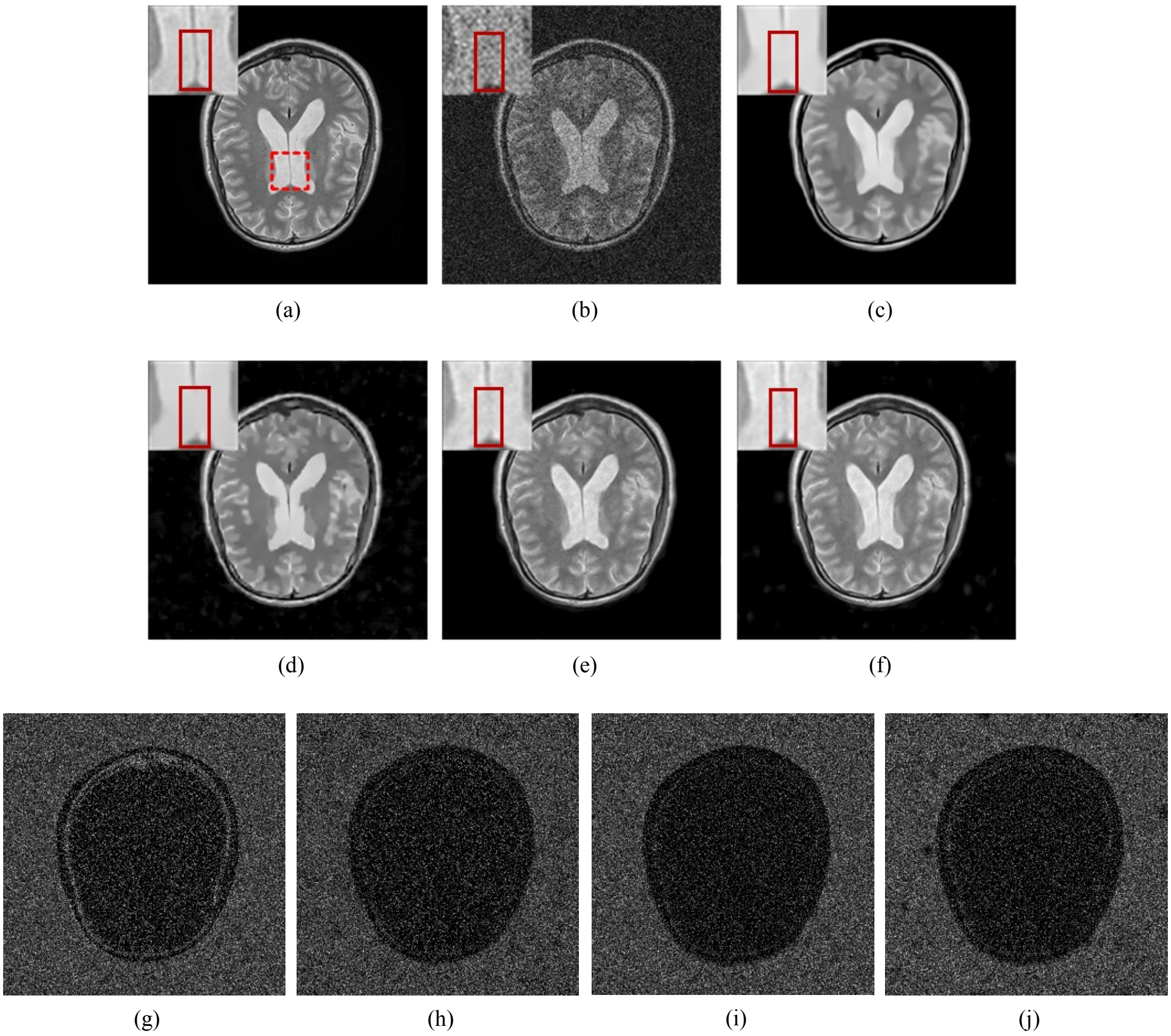

(a) (b) (c) (d) (e) (f) (g) (h) (i) (j)

**Fig. 9.** One denoised PDw example from the Guy's Hospital dataset with 15% Rician noise: (a) Noise-free image, (b) Noisy image, (c) BM4D, (d) PRI-NLM3D, (e) CNN3D, (f) RED-WGAN, (g) residual of BM4D, (h) residual of PRI-NLM3D, (i) residual of CNN3D, (j) residual of RED-WGAN.

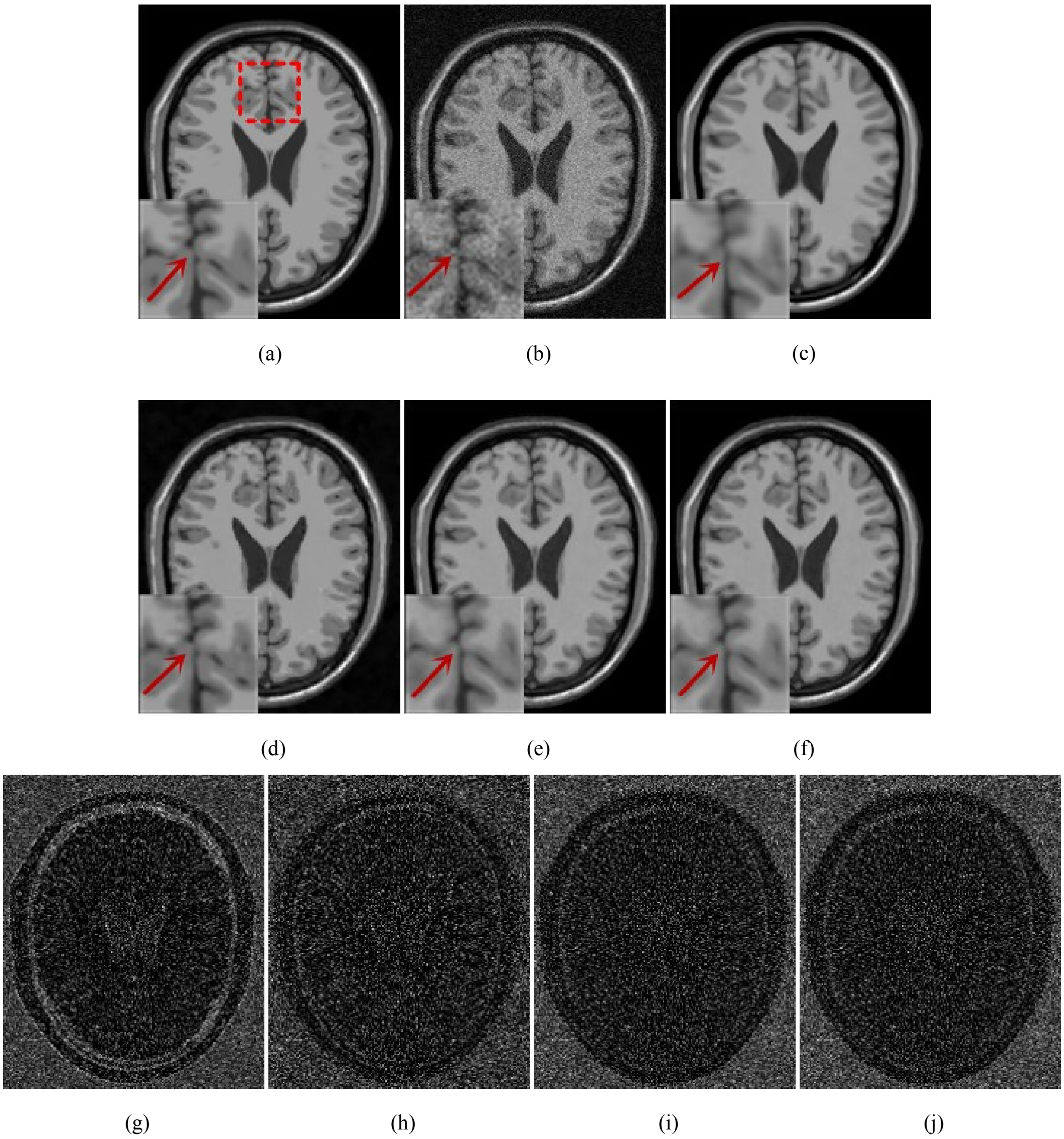

Figure panel labels: (a) (b) (c) (d) (e) (f) (g) (h) (i) (j)

**Fig. 10.** One denoised T1w example from the BrainWeb dataset with 9% Rician noise. (a) Noise-free image, (b) Noisy image (PSNR =24.0697, SSIM=0.5847, IFC=2.8367), (c) BM4D (PSNR=30.2730, SSIM=0.9335, IFC=3.3105), (d) PRI-NLM3D (PSNR=32.4518, SSIM=0.8 099, IFC=3.6819), (e) CNN3D (PSNR=34.2789, SSIM=0.9665, IFC=4.0195), (f) RED-WGAN (**PSNR=34.7432, SSIM=0.9706, IFC=4.1887**) (g) Residual of BM4D, (h) Residual of PRI-NLM3D, (i) Residual of CNN3D, (j) Residual of RED-WGAN.

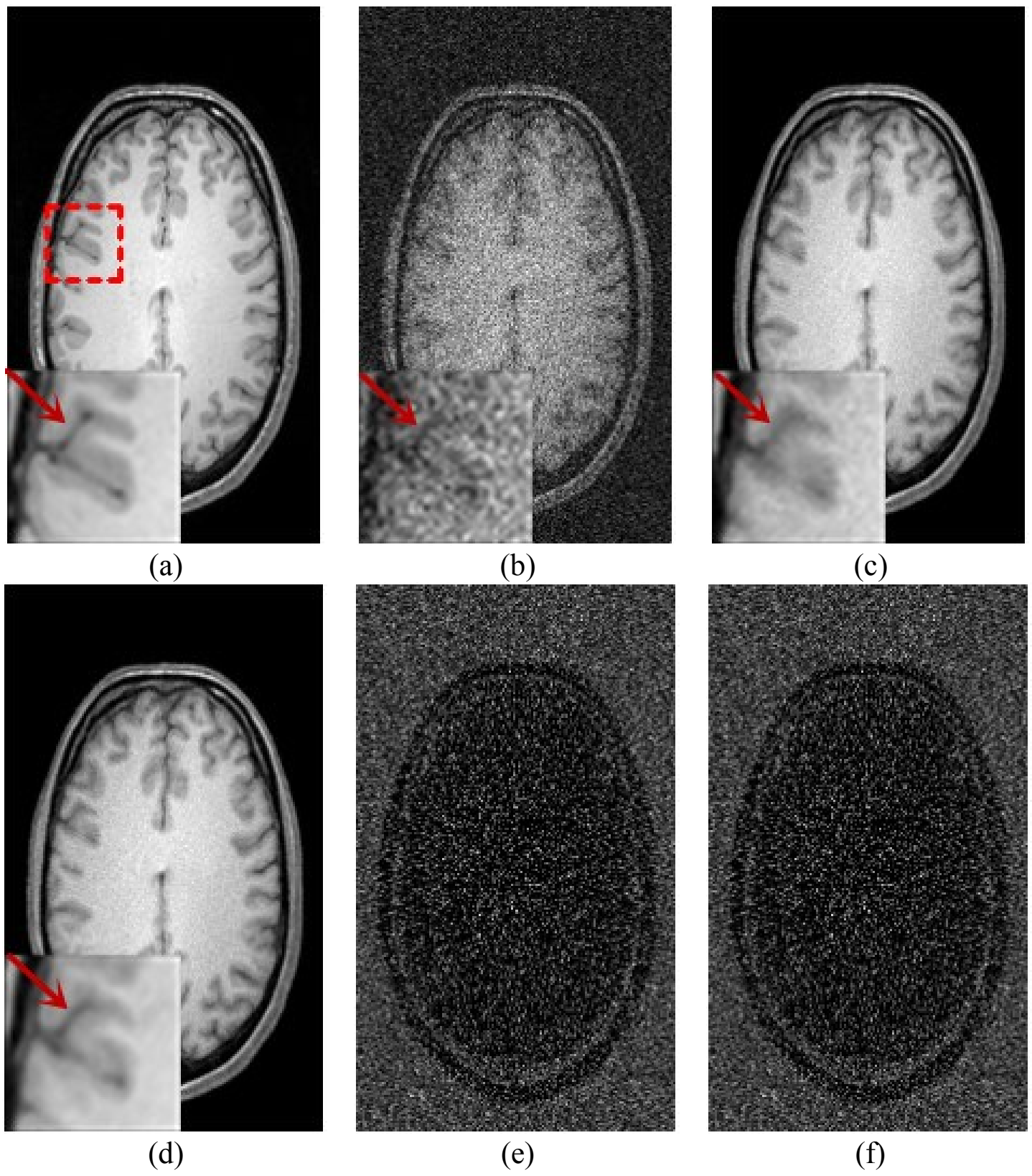

**Fig. 11.** One denoised T1w example from the testing set with 15% Rician noise. (a) Noise-free image, (b) Noisy image, (c) WGAN-MSE, (d) RED-WGAN, (e) Residual of WGAN-MSE, (f) Residual of RED-WGAN.

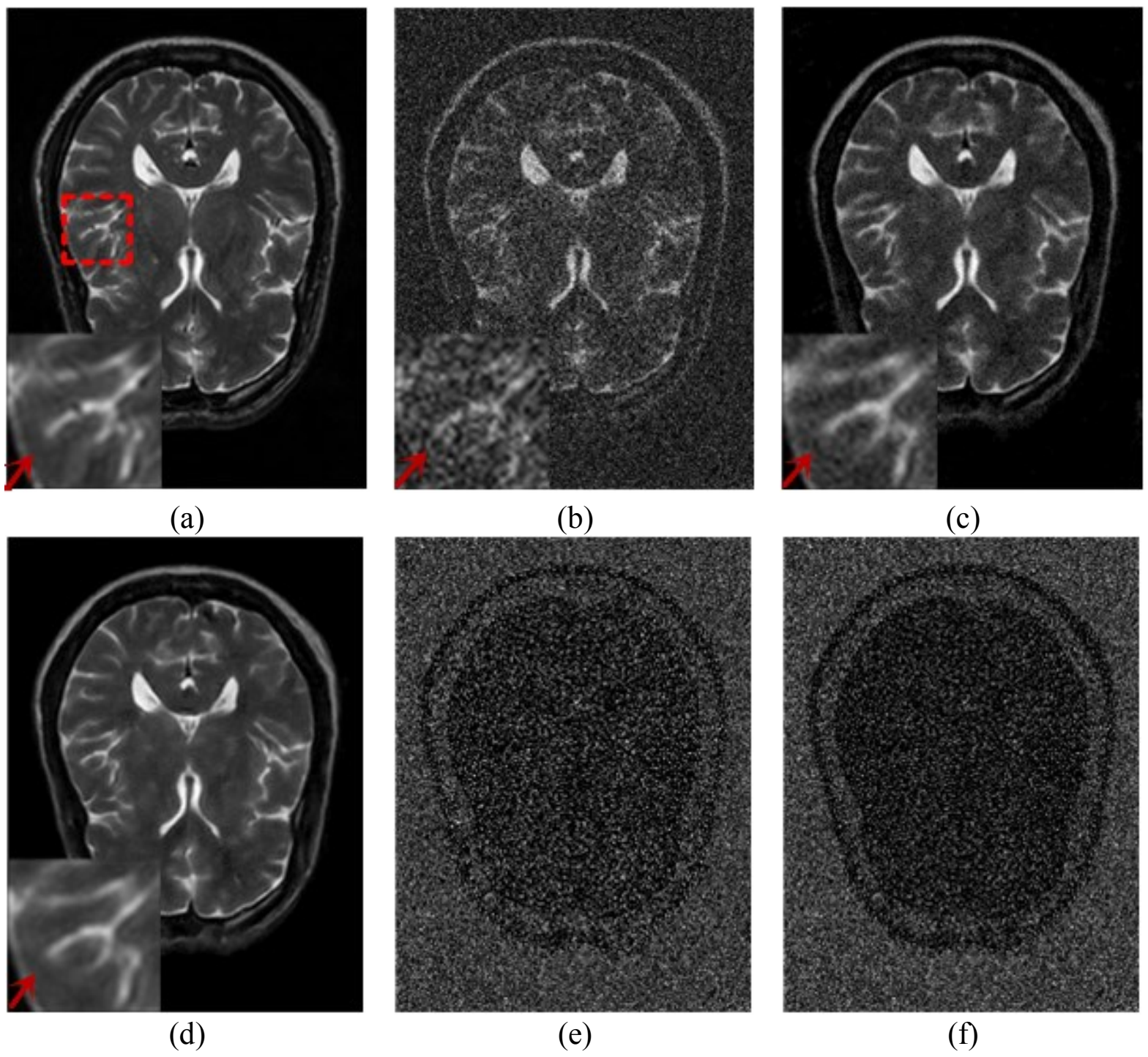

**Fig. 12.** One denoised T2w example from the testing set with 15% Rician noise. (a) Noise-free image, (b) Noisy image, (c) WGAN-MSE, (d) RED-WGAN, (e) Residual of WGAN-MSE, (f) Residual of RED-WGAN.

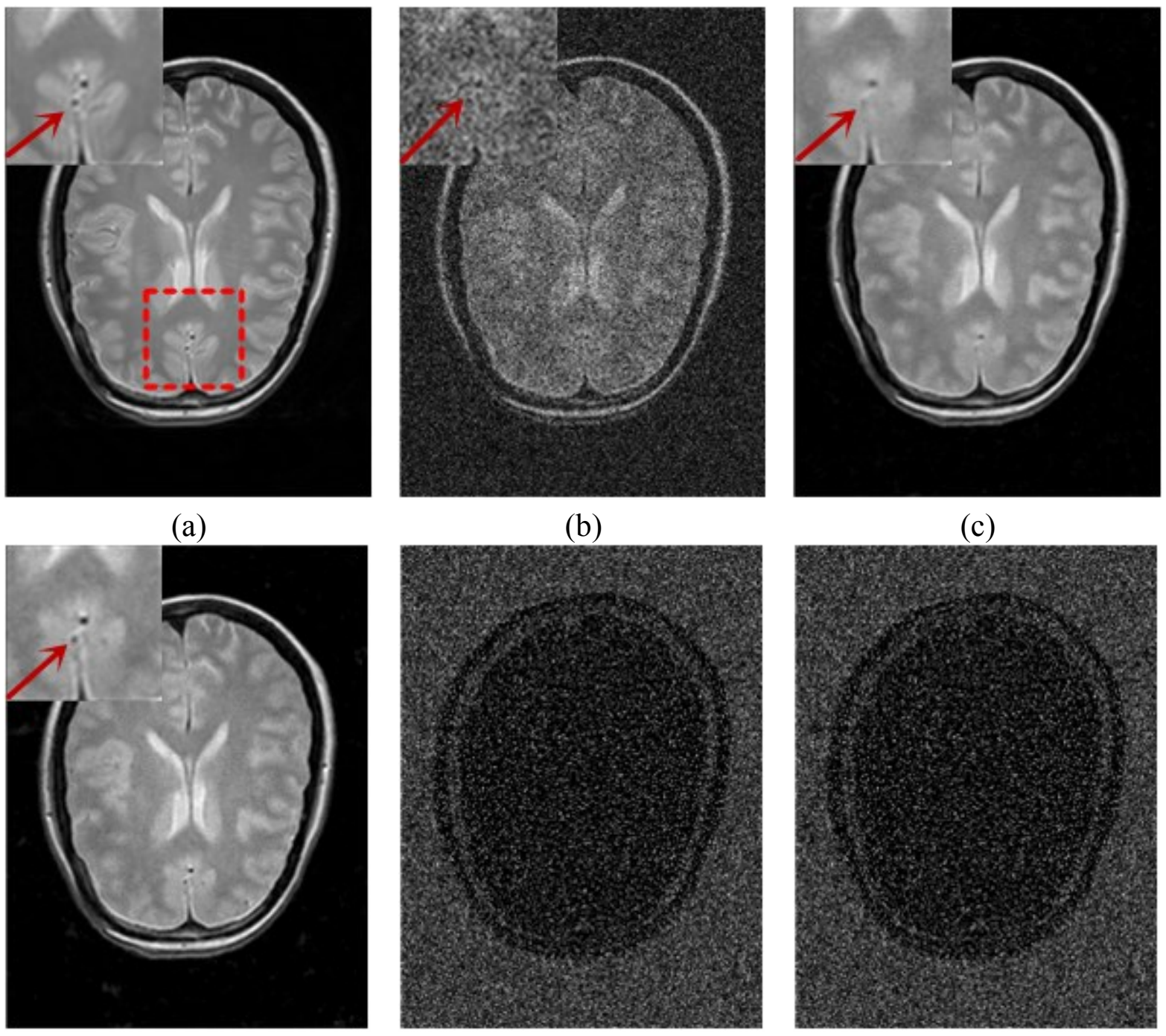

(a) (b) (c) (d) (e) (f)

**Fig. 13.** One denoised PDw example from the testing set with 15% Rician noise. (a) Noise-free image, (b) Noisy image, (c) WGAN-MSE, (d) RED-WGAN, (e) Residual of WGAN-MSE, (f) Residual of RED-WGAN.

**Table 6**

Quantitative results associated with different method outputs in Figs. 11 (T1w), 12 (T2w) and 13 (PDw).

| Method | T1w | | | T2w | | | PDw | | |
|---|---|---|---|---|---|---|---|---|---|
| | PSNR | SSIM | IFC | PSNR | SSIM | IFC | PSNR | SSIM | IFC |
| Noise | 15.0129 | 0.2698 | 1.2038 | 15.0275 | 0.2108 | 1.0588 | 14.6892 | 0.1966 | 1.0953 |
| WGAN-MSE | 28.9253 | 0.8280 | 2.1799 | 26.7390 | 0.7212 | 1.8081 | 31.7060 | **0.8449** | 2.2641 |
| RED-WGAN | **29.8372** | **0.8743** | **2.2834** | **29.3393** | **0.8156** | **1.9494** | **32.2121** | 0.8309 | **2.4691** |

### 3）Impact of perceptual loss

In this section, to sense the impact of perceptual loss, we compare RED-WGAN with the variant that has the same structure as RED-WGAN but with only the MSE loss, and we denote it as WGAN-MSE. Figs. 11-13 show examples of denoising results with both RED-WGAN and WGAN-MSE on T1w, T2w and PDw images, respectively. From the results, it can be observed that the noise in WGAN-MSE is not well suppressed, and some structures are blurred. On the other hand, RED-WGAN eliminates most of the noise and preserves the structural details better than WGAN-MSE. In Fig. 13, the blood vessel marked by the red arrows is very hard to identify, but RED-WGAN preserves it. Table 6 shows the quantitative results for Figs. 11-13. Table 6 shows that RED-WGAN outperforms WGAN-MSE in most cases, which is consistent with the visual results.

### 3.5 Robustness Analysis

For the results shown in the previous section, the proposed RED-WGAN model was trained and tested with images with identical noise levels, which is a difficult requirement to meet in practice. Meanwhile, although the proposed model was validated by different datasets with various scanners and scanning parameters, the noise in these images was simulated, which may be different from real situations. To validate the robustness of the proposed model under these two scenarios, additional experiments were performed. All the models used in this section were also trained with the Hammersmith dataset.

#### 1) Comparisons with the Special Model on Various Noise Levels

To show the robustness of RED-WGAN for various noise levels, three RED-WGAN models were trained with different noise levels: 1%, 9% and 15%. We denote these models as RED-WGAN-n, where n indicates the corresponding noise level. Meanwhile, the same model denoted as RED-WGAN-m was trained with a mix of noise levels, which ranged from 1% to 19% with a step of 2%. These four models were tested on the images from the Hammersmith dataset with various noise levels. The quantitative results are shown in Table 7. We denote the best results in red and the second best in blue. It is clear that RED-WGAN-m is superior to the other methods in most situations, which means that for real clinical applications without prior knowledge about the noise level, training the DL model with a mix of possible noise levels is one of the potential solutions. The performance of RED-WGAN-m is slightly worse than traditional methods at a low noise level (1%). The possible reason is that when simultaneously training with higher noise levels, the risk that the network may mistreat the noise as details from a low noise level is increased.

It also can be observed that the model RED-WGAN-n trained with a single noise level of n% can efficiently cover a certain noise range. For example, RED-WGAN-9, which was trained with a 9% noise level, has better scores on the testing set with 7% to 13% noise levels. This can also be seen as solid evidence for the generalization and robustness of our model, as most traditional methods also need to adjust the parameters to fit the different noise levels.

#### 2) Real MR data

The propose of this subsection is to verify the effectiveness of the proposed model on real noisy clinical data. The experiments were conducted on two brain MR image volumes, which belong to a human being and a mouse, respectively. The human brain image was acquired on a Siemens (Erlangen, Germany) Trio Tim 3T scanner using an MP-RAGE sequence with TR=2400 ms, TE=2.01 ms, TI=1000 ms, flip angle=8, voxel resolution=0.8×0.8×0.8 mm3 and 256×256×224 voxels. The mouse brain image was acquired on a Bruker BioSpec 7T scanner using a 3D RARE sequence with a TR=1200, an effective TE=62.5 ms, a RARE factor=16, a voxel resolution=0.1×0.1×0.1 mm3 and 225×192×96 voxels.

Due to the lack of knowledge about the noise level in the real data, we experimentally selected RED-WGAN models trained with 1% and 4% noise for human and mouse data, respectively. Since ground truth images are unavailable, the SNR was measured in a homogeneous region and used as the quantitative metric. The results are shown in Figs. 14 and 15. In Fig. 14, it is clear that the traditional methods cannot eliminate all the noise in the brain, especially in the epencephalon and brainstem, but RED-WGAN can efficiently suppress most of the noise, even near the epencephalon and brainstem, which are indicated by red arrows. It is noted that a certain level of noises in homogeneous areas can be noticed in Fig. 14(d). In Fig. 15, the noise is much heavier than Fig. 14. All the methods can remove most of the noise, but the results of BM4D and PRI-NLM3D look oversmoothed, and RED-WGAN obtained better visual effects and preserved more details. Furthermore, RED-WGAN obtained a better SNR in both cases.

**Table 7**

From top to bottom, the PSNR, SSIM and IFC measures of special methods on T1w images with different noise levels

| | 1% | 3% | 5% | 7% | 9% | 11% | 13% | 15% | 17% | 19% |
|---|---|---|---|---|---|---|---|---|---|---|
| Noise | 39.2102 | 29.222 | 24.6338 | 21.6303 | 19.3947 | 17.6232 | 16.1494 | 14.8897 | 13.7963 | 12.8121 |
| | 0.8324 | 0.6008 | 0.4965 | 0.4241 | 0.3662 | 0.3181 | 0.2771 | 0.2417 | 0.2114 | 0.1849 |
| | 6.9754 | 3.8463 | 2.7421 | 2.1253 | 1.7185 | 1.4412 | 1.2245 | 1.0575 | 0.9228 | 0.8107 |
| PRI-NLM3D | 42.3523 | **36.6855** | 33.8387 | 31.4226 | 29.9465 | 28.3322 | 27.4751 | 25.9823 | 25.5629 | 24.0545 |
| | 0.9597 | 0.9211 | 0.8906 | 0.8116 | 0.7784 | 0.7244 | 0.6940 | 0.6529 | 0.6540 | 0.5917 |
| | 6.8742 | 4.4655 | 3.4645 | 2.8561 | 2.4292 | 2.1327 | 1.8660 | 1.6781 | 1.5103 | 1.3814 |
| BM4D | **43.7217** | **37.3037** | **34.5095** | **32.6762** | 31.3338 | 29.7973 | 28.1597 | 25.9018 | 23.1987 | 21.1881 |
| | **0.9832** | **0.9393** | **0.9034** | **0.8926** | 0.8798 | **0.8622** | 0.8417 | 0.8126 | **0.7727** | **0.7320** |
| | 7.3469 | **4.6026** | **3.6226** | 3.0635 | 2.6889 | 2.3640 | 2.0850 | 1.8054 | 1.5234 | 1.3300 |
| RED-WGAN-1 | **44.5211** | 32.0828 | 26.1881 | 22.6619 | 20.1564 | 18.227 | 16.6428 | 15.3077 | 14.1590 | 13.1323 |
| | **0.9804** | 0.6716 | 0.5358 | 0.4527 | 0.3893 | 0.3375 | 0.2936 | 0.2560 | 0.2236 | 0.1954 |
| | **7.5668** | 4.1701 | 2.9079 | 2.2258 | 1.7858 | 1.4899 | 1.2609 | 1.0856 | 0.9452 | 0.8287 |
| RED-WGAN-9 | 31.2806 | 31.4654 | 31.7522 | 32.2856 | **33.0640** | **31.1793** | 28.1836 | 24.9473 | 22.1897 | 19.9061 |
| | 0.8746 | 0.8711 | 0.8726 | 0.8828 | **0.9009** | 0.8392 | 0.6572 | 0.5206 | 0.4296 | 0.3664 |
| | 2.9934 | 3.0662 | 3.1033 | **3.1022** | **3.1470** | **2.6533** | 2.2611 | 1.8946 | 1.5957 | 1.3478 |
| RED-WGAN-15 | 25.5915 | 26.0521 | 26.5919 | 27.2255 | 27.9520 | 28.7735 | **29.5500** | **29.6708** | **28.3798** | **25.7872** |
| | 0.8041 | 0.7978 | 0.7946 | 0.7996 | 0.8108 | 0.8276 | **0.8446** | **0.8212** | 0.7129 | 0.5747 |
| | 1.8390 | 2.0015 | 2.0015 | 2.1009 | 2.1997 | 2.2984 | **2.3404** | **2.2839** | **2.0773** | **1.8279** |
| RED-WGAN-m | 41.5200 | 36.1271 | **33.9343** | **33.3728** | **32.8044** | **32.4096** | **31.9193** | **31.1414** | **29.9954** | **28.2576** |
| | 0.9495 | **0.9298** | **0.9053** | **0.9066** | **0.9031** | **0.8995** | **0.8958** | **0.8849** | **0.8556** | **0.8044** |
| | **6.8936** | **4.4793** | **3.6446** | **3.3569** | **3.1052** | **2.9320** | **2.7547** | **2.5607** | **2.3460** | **2.0864** |

## 4 Discussions and Conclusions

In this paper, we propose a novel method based on the Wasserstein generative adversarial network to remove the Rician noise in MR images while effectively preserving the structural details. This network aims to process 3D volume data using a 3D convolutional neural network. In addition to the introduction of the WGAN framework, there are two more advantages to our method: the innovative generator structure and mixed weighted loss function. The generator is constructed with an autoencoder structure, which symmetrically contains convolutional and deconvolutional layers, aided by a residual structure. Another improvement of our method is the adaptation of the mixed loss function, which combines the MSE and perceptual losses with a weighted form.

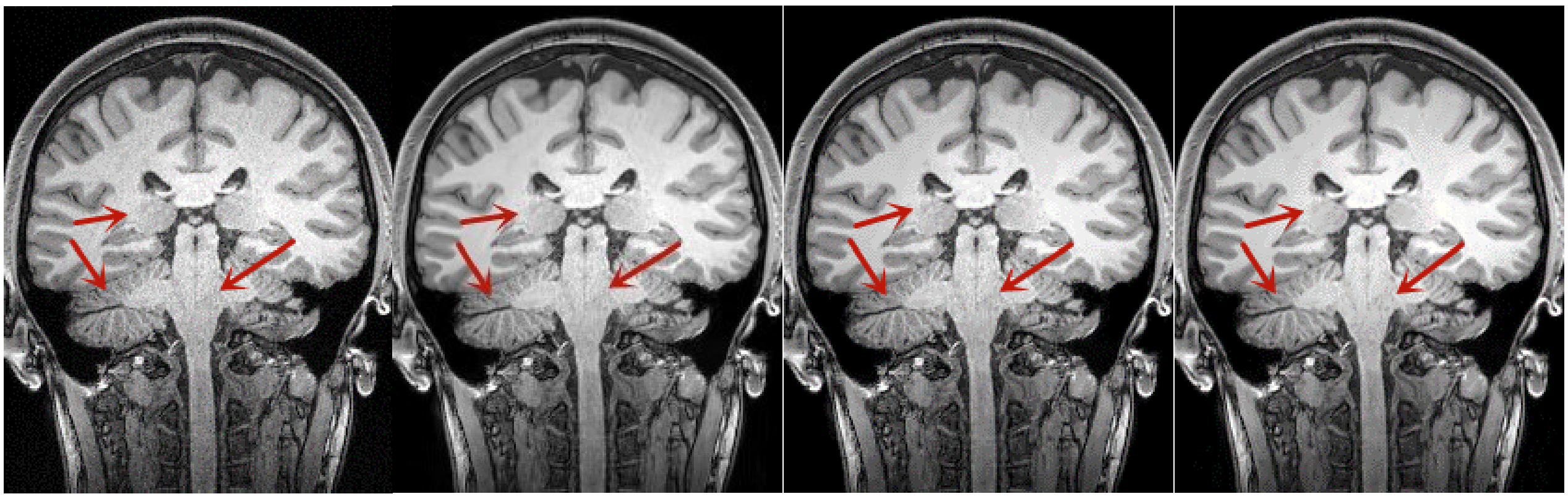

(a)Noisy (SNR=26.52 dB) | (b)BM4D (SNR=27.41 dB) | (c)PRI-NLM3D (SNR=27.00 dB) | (d) RED-WGAN (**SNR=28.34 dB**)

**Fig. 14.** Denoised result on real T1w human brain data.

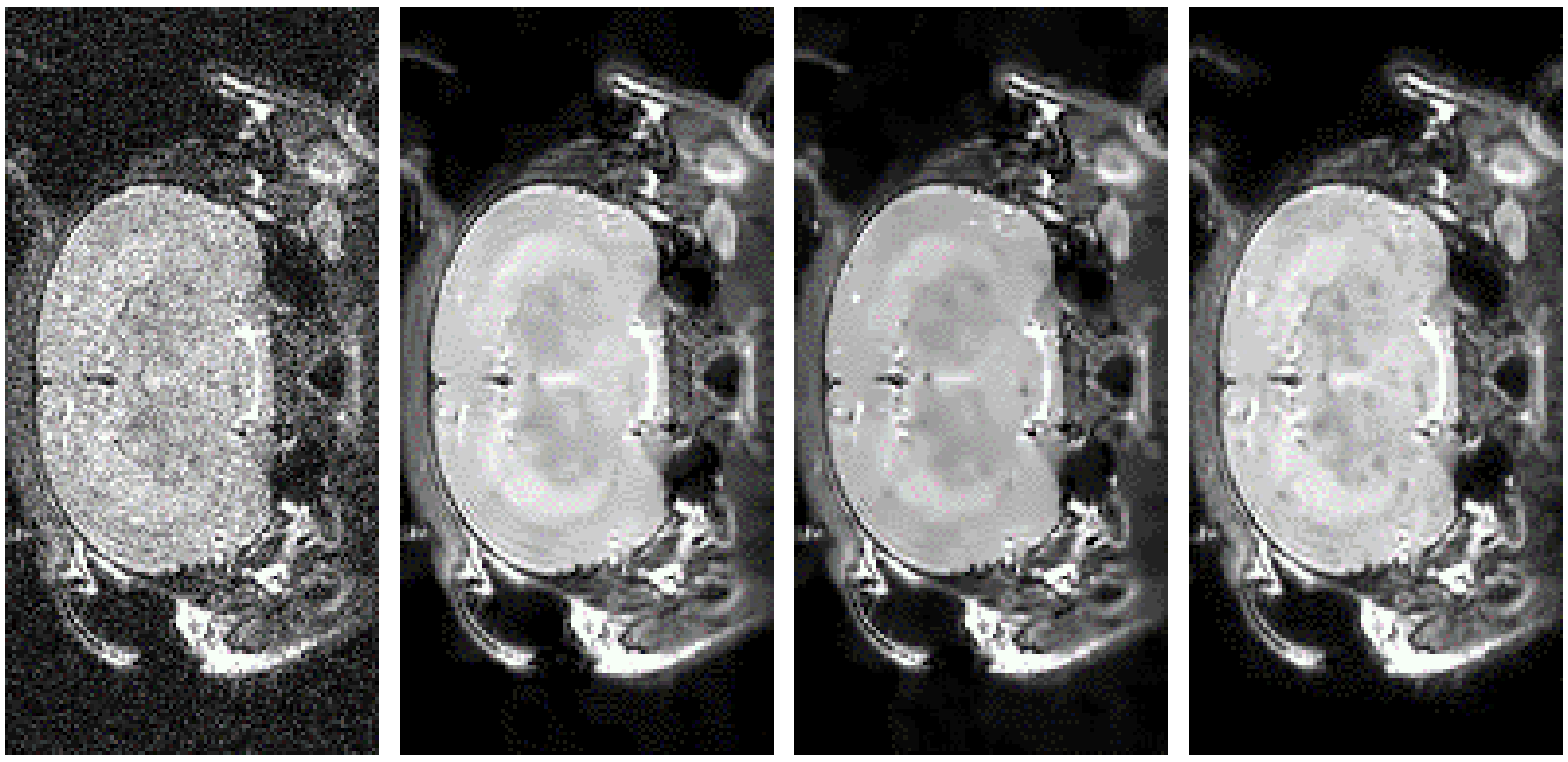

(a) Noisy (SNR=30.38 dB) | (b)BM4D (SNR=37.62 dB) | (c)PRI-NLM3D (SNR=39.00 dB) | (d) RED-WGAN (**SNR=39.41 dB**)

**Fig. 15.** Denoised result on real T1w mouse brain data.

The experimental results demonstrate that with the help of WGAN and perceptual loss, the CNN-based method is significantly improved in both qualitative and quantitative aspects. Compared to several state-of-the-art methods, including BM3D, PRI-NLM3D and CNN3D, our proposed RED-WGAN effectively avoids oversmoothing effects while preserving more details. Furthermore, to validate the robustness and generalization of our model, we trained our model with several specific noise levels and tested it on various noise levels. Meanwhile, real noisy clinical data were involved. In both cases, the proposed RED-WGAN model achieved a performance better than the traditional methods in both visual effects and quantitative results.

The computational cost of the deep learning-based method is worth mentioning. The training stage is the costliest step. Although the training procedure is usually performed on the GPU, it is still time consuming. For our training set, when we alternatingly train the generator and discriminator networks, each epoch takes approximately 40 minutes. Although other methods, such as BM4D and PRI-NLM3D, do not need to train, their running times are much longer than the DL-based methods. In this paper, the average execution times for the clinical dataset for BM4D, PRI-NLM3D, CNN3D and RED-WGAN were 5.73, 4.16, 0.17 and 0.16 s, respectively. In practice, the running time for DL-based methods can be further reduced by using GPU for testing.

In conclusion, the results obtained in the paper are encouraging and efficiently demonstrate the potential of deep learning-based methods for MRI denoising. In the future, instead of training on a specific noise level, we will try to extend our method to a more general form for different noise levels. Furthermore, incorporating the image reconstruction method may be interesting.

**Acknowledgments**

This work was supported in part by the National Natural Science Foundation of China under grants 61671312, 61871277, 81530060 and 81471752, the National Key R&D Program of China under grant 2017YFB0802300, 2017YFA0104302, 2017YFC0109202 and 2017YFC0107900, the Science and Technology Project of Sichuan Province of China under grant 2018HH0070, and the Guangdong Provincial Key Laboratory of Medical Image Processing (2017B030314133).

**References**

Anand, C.S., Sahambi, J.S., 2009. MRI denoising using bilateral filter in redundant wavelet domain, In: Proc. IEEE Region Conf.(TENCON ) 2009, pp. 1-6.

Andersen, A.H., 1996. On the Rician distribution of noisy MRI data. Magn. Reson. Med. 36, 331-333.

Arjovsky, M., Bottou, L., 2017. Towards principled methods for training generative adversarial networks. arXiv:1701.04862.

Arjovsky, M., Chintala, S., Bottou, L., 2017. Wasserstein GAN. arXiv preprint arXiv:1701.07875, 2017.

Awate, S.P., Whitaker, R.T., 2007. Feature-preserving MRI denoising: a nonparametric empirical Bayes approach. IEEE Trans. Med. Imaging 26, 1242-1255.

Bouhrara, M., Bonny, J.M., Ashinsky, B., Maring, M., Spencer, R., 2016. Noise estimation and reduction in multispectral

magnetic resonance images. IEEE Trans. Med. Imaging 36, 181-193.

Bruna, J., Sprechmann, P., Lecun, Y., 2015. Super-resolution with deep convolutional sufficient statistics, In: Proc. Int. Conf. Learning Representations （ICLR) 2015.

Buades, A., Coll, B., Morel, J.M., 2005. A non-local algorithm for image denoising, In: Proc. IEEE Conf. Comp. Vis. Patt. Recogn. (CVPR) 2005, pp. 60-65

Chen, H., Zhang, Y., Kalra, M.K., Lin, F., Chen, Y., Liao, P., Zhou, J., Wang, G., 2017a. Low-dose CT with a residual encoder-decoder convolutional neural network. IEEE Trans. Med. Imaging 36, 2524-2535.

Chen, H., Zhang, Y., Zhang, W., Liao, P., Li, K., Zhou, J., Wang, G., 2017b. Low-dose CT via convolutional neural network. Biomed. Opt. Express 8, 679-694.

Coupé, P., Yger, P., Prima, S., Hellier, P., Kervrann, C., Barillot, C., 2008. An optimized blockwise nonlocal means denoising filter for 3-D magnetic resonance images. IEEE Trans. Med. Imaging 27, 425-441.

Dabov, K., Foi, A., Katkovnik, V., Egiazarian, K., 2007. Image denoising by sparse 3-D transform-domain collaborative filtering. IEEE Trans. Image Process. 16, 2080-2095.

Dong, C., Chen, C.L., He, K., Tang, X., 2014. Learning a deep convolutional network for image super-resolution, In: Proc. 13th Eur. Conf. Comput. Vis.(ECCV) 2014, pp. 184-199.

Dong, C., Chen, C.L., He, K., Tang, X., 2016. Image super-resolution using deep convolutional networks. IEEE Trans. Pattern Anal. Mach. Intell. 38, 295-307.

Xie, J., Xu, L., Chen, E., 2012. Image denoising and inpainting with deep neural networks, In: Proc. Adv. Neural Inf. Process. Syst.(NIPS) 2012, pp. 341-349.

Gatys, L.A., Ecker, A.S., Bethge, M., 2015. Texture synthesis using convolutional neural networks, In: Proc. Adv. Neural Inf. Process. Syst. (NIPS), 2015, pp. 262-270.

Gerig, G., Kübler, O., Kikinis, R., Jolesz, F.A., 1992. Nonlinear anisotropic filtering of MRI data. IEEE Trans. Med. Imaging 11, 221-232.

Girshick, R., 2015. Fast R-CNN, In: Proc. IEEE Int. Conf. Comp. Vis. (ICCV), 2015, pp. pp. 1440-1448.

Golshan, H.M., Hasanzadeh, R.P., Yousefzadeh, S.C., 2013. An MRI denoising method using image data redundancy and local SNR estimation. Magn. Reson. Imaging 31, 1206-1217.

Goodfellow, I.J., Pouget-Abadie, J., Mirza, M., Xu, B., Warde-Farley, D., Ozair, S., Courville, A., Bengio, Y., 2014. Generative adversarial nets, In: Proc. Adv. Neural Inf. Process. Syst.(NIPS) 2014, pp. 2672-2680.

Gulrajani, I., Ahmed, F., Arjovsky, M., Dumoulin, V., Courville, A., 2017. Improved training of Wasserstein GANs, In:

Proc. Adv. Neural Inf. Process. Syst. (NIPS) 2017, pp. 5767-5777.

Hamid Rahim, S., Alan Conrad, B., Gustavo, D.V., 2005. An information fidelity criterion for image quality assessment using natural scene statistics. IEEE Trans. Image Process. 14, 2117-2128.

He, K., Gkioxari, G., Dollár, P., Girshick, R., 2017. Mask R-CNN, In: Proc. IEEE Int. Conf. Comp. Vis.(ICCV) 2017, pp. 2980-2988.

He, K., Zhang, X., Ren, S., Sun, J., 2016. Deep residual learning for image recognition, In: Proc. IEEE Conf. Comp. Vis. Patt. Recogn. (CVPR) 2016, pp. 770-778.

He, L., Greenshields, I.R., 2009. A nonlocal maximum likelihood estimation method for Rician noise reduction in MR images. IEEE Trans. Med. Imaging 28, 165-172.

Hu, J., Pu, Y., Wu, X., Zhang, Y., Zhou, J., 2012. Improved DCT-based nonlocal means filter for MR images denoising. Comput. Math. Method Med. 2012.

Hu, J., Zhou, J., Wu, X., 2016. Non-local MRI denoising using random sampling. Magn. Reson. Imaging 34, 990-999.

Isola, P., Zhu, J.Y., Zhou, T., Efros, A.A., 2017. Image-to-image translation with conditional adversarial networks, In: Proc. IEEE Conf. Comp. Vis. Patt. Recogn.(CVPR) 2017, pp. 1125-1134.

Jain, V., Seung, H.S., 2008. Natural image denoising with convolutional networks, In: Proc. Adv. Neural Inf. Process. Syst. (NIPS) 2008, pp. 769-776.

Jiang, D., Dou, W., Vosters, L., Xu, X., Sun, Y., Tan, T., 2018. Denoising of 3D magnetic resonance images with multi-channel residual learning of convolutional neural network. Jpn. J. Radiol. 36, pp 566-574.

Johnson, J., Alahi, A., Li, F.F., 2016. Perceptual losses for real-time style transfer and super-resolution. Proc. Eur. Conf. Comput. Vis. (ECCV), 694-711.

Kataoka, Y., Matsubara, T., Uehara, K., 2016. Image generation using generative adversarial networks and attention mechanism, In: Proc. IEEE/ACIS 15th Int. Conf. Comput. Inf. Sci. (ICIS) 2016, pp. 1-6.

Kinga, D., Adam, J.B., 2015. Adam: A method for stochastic optimization, In: Proc. Int. Conf. Learning Representations (ICLR) 2015.

Krissian, K., Aja-Fernandez, S., 2009. Noise-driven anisotropic diffusion filtering of MRI. IEEE Trans.Image Process. 18, 2265-2274.

Ledig, C., Theis, L., Huszár, F., Caballero, J., Cunningham, A., Acosta, A., Aitken, A.P., Tejani, A., Totz, J., Wang, Z., 2017. Photo-realistic single image super-resolution using a generative adversarial network, In: Proc. IEEE Conf. Comp. Vis. Patt. Recogn. (CVPR) 2017, pp. 4681-4690.

Li, H., Mueller, K., 2017. Low-dose CT streak artifacts removal using deep residual neural network, In: Proc. Fully Three-Dimensional Image Reconstruction Radiol. Nucl. Med. (Fully3D) 2017, pp. 191-194.

Li, R., Zhang, W., Suk, H.I., Wang, L., Li, J., Shen, D., Ji, S., 2014. Deep learning based imaging data completion for improved brain disease diagnosis, In: Int. Conf. Med. Image Comput. Comput. Assist. Interv.(MICCAI) 2014, pp. 305-312.

Liu, Y., Zhang, Y., 2018. Low-dose CT restoration via stacked sparse denoising autoencoders. Neurocomputing 284, 80-89.

Long, J., Shelhamer, E., Darrell, T., 2015. Fully convolutional networks for semantic segmentation, In: Proc. IEEE Conf. Comp. Vis. Patt. Recogn. (CVPR) 2015, pp. 3431-3440.

Ma, J., Plonka, G., 2007. Combined curvelet shrinkage and nonlinear anisotropic diffusion. IEEE Trans. Image Process. 16, 2198-2206.

Maggioni, M., Katkovnik, V., Egiazarian, K., Foi, A., 2012. Nonlocal transform-domain filter for volumetric data denoising and reconstruction. IEEE Trans. Image Process. 22, 119-133.

Manjon, J., Carbonell-Caballero, J., Jj, Garcia-Marti, G., Marti-Bonmati, L., Robles, M., 2008. MRI denoising using non-local means. Med. Image Anal. 12, 514-523.

Manjón, J.V., Coupé, P., Buades, A., Louis, C.D., Robles, M., 2012. New methods for MRI denoising based on sparseness and self-similarity. Med. Image Anal. 16, 18-27.

Mcveigh, E.R., Henkelman, R.M., Bronskill, M.J., 1985. Noise and filtration in magnetic resonance imaging. Med. Phys. 12, 586-591.

Mohan, J., Krishnaveni, V., Guo, Y., 2014. A survey on the magnetic resonance image denoising methods. Biomed. Signal Process. Control 9, 56-69.

Nowak, R.D., 1999. Wavelet-based Rician noise removal for magnetic resonance imaging. IEEE Trans. Image Process. 8, 1408-1419.

Pal, C., Das, P., Chakrabarti, A., Ghosh, R., 2017. Rician noise removal in magnitude MRI images using efficient anisotropic diffusion filtering. Int. J. Imaging Syst. Technol. 27, 248-264.

Pan, S.J., Yang, Q., 2010. A survey on transfer learning. IEEE Trans. Knowl. Data Eng. 22, 1345-1359.

Perona, P., Malik, J., 1990. Scale-space and edge detection using anisotropic diffusion. IEEE Trans. Pattern Anal. Mach. Intell.

Pizurica, A., Philips, W., Lemahieu, I., Acheroy, M., 2003. A versatile wavelet domain noise filtration technique for medical imaging. IEEE Trans. Med. Imaging 22, 323-331.

Rajan, J., Veraart, J., Van Audekerke, J., Verhoye, M., Sijbers, J., 2012. Nonlocal maximum likelihood estimation method for denoising multiple-coil magnetic resonance images. Magn. Reson. Imaging 30, 1512-1518.

Rajwade, A., Rangarajan, A., Banerjee, A., 2013. Image denoising using the higher order singular value decomposition. IEEE Trans. Pattern Anal. Mach. Intell. 35, 849-862.

Ronneberger, O., Fischer, P., Brox, T., 2015. U-net: Convolutional networks for biomedical image segmentation, In: Int. Conf. Med. Image Comput. Comput. Assist. Interv. (MICCAI) 2015, pp. 234-241.

Samsonov, A.A., Johnson, C.R., 2004. Noise-adaptive nonlinear diffusion filtering of MR images with spatially varying noise levels. Magn. Reson. Med. 52, 798-806.

Shan, H., Zhang, Y., Yang, Q., Kruger, U., Kalra, M.K., Sun, L., Cong, W., Wang, G., 2018. 3-D convolutional encoder-decoder network for low-dose CT via transfer learning from a 2-D trained network. IEEE Trans. Med. Imaging 37, 1522-1534.

Sijbers, J., den Dekker, A.J., Scheunders, P., Van, D.D., 1998. Maximum-likelihood estimation of Rician distribution parameters. IEEE Trans. Med. Imaging 17, 357-361.

Simonyan, K., Zisserman, A., 2014. Very deep convolutional networks for large-scale image recognition. Computer Science.

Vincent, P., Larochelle, H., Lajoie, I., Bengio, Y., Manzagol, P.A., 2010. Stacked denoising autoencoders: learning useful representations in a deep network with a local denoising criterion. J. Mach. Learn. Res. 11, 3371-3408.

Wang, S., Su, Z., Ying, L., Peng, X., Zhu, S., Liang, F., Feng, D., Liang, D., 2016. Accelerating magnetic resonance imaging via deep learning, In: Proc. IEEE Int. Symp. Biomed. Imaging 2016, pp. 514-517.

Wang, Z., Bovik, A.C., Sheikh, H.R., Simoncelli, E.P., 2004. Image quality assessment: from error measurement to structural similarity. IEEE Trans. Image Process. 13, 600-612.

Wiest-Daesslé, N., Prima, S., Coupé, P., Morrissey, S.P., Barillot, C., 2008. Rician noise removal by non-local means filtering for low signal-to-noise ratio MRI: applications to DT-MRI, In: Int. Conf. Med. Image Comput. Comput. Assist. Interv. (MICCAI) 2008, pp. 171-179.

Xiang, L., Qiao, Y., Nie, D., An, L., Wang, Q., Shen, D., 2017. Deep auto-context convolutional neural networks for standard-dose pet image estimation from low-dose PET/MRI. Neurocomputing 267, 406-416.

Xu, J., Gong, E., Pauly, J., Zaharchuk, G., 2017. 200x low-dose PET reconstruction using deep learning. arXiv preprint arXiv:1712.04119.

Yang, G., Yu, S., Dong, H., Slabaugh, G., Dragotti, P.L., Ye, X., Liu, F., Arridge, S., Keegan, J., Guo, Y., 2018. DAGAN: deep de-aliasing generative adversarial networks for fast compressed sensing MRI reconstruction. IEEE Trans. Med. Imaging 37, 1310-1321.

Yang, Q., Yan, P., Zhang, Y., Yu, H., Shi, Y., Mou, X., Kalra, M.K., Zhang, Y., Sun, L., Wang, G., 2017a. Low-dose CT image denoising using a generative adversarial network with wasserstein distance and perceptual loss. IEEE Trans. Med. Imaging 37, 1348-1357.

Yang, W., Zhang, H., Yang, J., Wu, J., Yin, X., Chen, Y., Shu, H., Luo, L., Coatrieux, G., Gui, Z., 2017b. Improving low-dose CT image using residual convolutional network. IEEE Access 5, 24698-24705.

Yaroslavsky, L.P., Egiazarian, K.O., Astola, J.T., 2001. Transform domain image restoration methods: review, comparison, and interpretation. Proc. SPIE, Nonlinear Image Process. Pattern Anal. XII, 155-169.

You, C., Zhang, Y., Zhang, X., Li, G., Ju, S., Zhao, Z., Zhang, Z., Cong, W., Saha, P.K., Wang, G., 2018. CT super-resolution gan constrained by the identical, residual, and cycle learning ensemble (GAN-CIRCLE). arXiv preprint arXiv:1808.04256.

Zhang, K., Zuo, W., Chen, Y., Meng, D., Zhang, L., 2017. Beyond a gaussian denoiser: residual learning of deep CNN for image denoising. IEEE Trans. Image Process. 26, 3142-3155.

Zhang, X., Xu, Z., Jia, N., Yang, W., Feng, Q., Chen, W., Feng, Y., 2015. Denoising of 3D magnetic resonance images by using higher-order singular value decomposition. Med. Image Anal. 19, 75-86.